\documentclass[aip,nofootinbib,12pt]{revtex4-1}
\usepackage{amssymb,amsmath,amsthm}
\usepackage{textcomp}

\usepackage{latexsym}
\usepackage[english]{babel} 
\usepackage[latin1]{inputenc}  
\usepackage[T1]{fontenc}   
\usepackage[all]{xy}
\usepackage{amsfonts}
\usepackage{graphics}
\usepackage{epsfig}
\usepackage{graphicx}
\usepackage[all]{xy}
\usepackage[colorlinks=true,linkcolor=blue]{hyperref}%
\usepackage{tikz}
\usetikzlibrary{automata}
\usetikzlibrary{arrows}
\usetikzlibrary{fit}
\usetikzlibrary{matrix}

\def\CC{\mathbb{C}}

\def\C{\mathbb{C}}

\def\QQ{\mathbb{Q}}
\def\PP{\mathbb{P}}

\newcommand{\incl}{\ar@{^{}-}}
\newcommand{\inclu}{\ar@{^{}.}}

\newtheorem{algo}{Algorithm}[section]

\newtheorem{lemma}{Lemma       }
\newtheorem{def }{Definition  }[section]
\newtheorem{theorem}{Theorem}
\newtheorem{ex}{Example     }[section]
\newtheorem{rem}{Remark    }[section]

\begin{document}

 \title{Entanglement of four qubit systems: a geometric atlas with polynomial compass  I (the finite world)}
\author{Fr\'ed\'eric Holweck\footnote{frederic.holweck@utbm.fr, 
Laboratoire IRTES-M3M,
Universit\'e de Technologie de Belfort-Montb\'eliard, 90010 Belfort Cedex, FR}, 
Jean-Gabriel Luque\footnote{jean-gabriel.luque@univ-rouen.fr, Universit\'e de Rouen, Laboratoire d'Informatique, du Traitement de l'Information et des Syst\`emes (LITIS), Avenue de l'Universit\'e - BP 8
6801 Saint-\'etienne-du-Rouvray Cedex, FR } and Jean-Yves Thibon\footnote{jyt@univ-mlv.fr, \ Labroatoire d'Informatique Gaspard Monge
Universit\'e Paris-Est Marne-la-Vall\'ee,
77454 Marne-la-Vall\'ee Cedex 2, FR}}

 \begin{abstract}
We investigate the geometry of the four qubit systems 
by means of algebraic geometry and invariant theory,
which allows us to interpret certain entangled
states as algebraic varieties. More precisely we describe  the nullcone, i.e., the 
set of states annihilated by all invariant polynomials,
and also the so called third secant variety, which
can be interpreted as the generalization of GHZ-states for more than three qubits. 
All our geometric descriptions go along
with algorithms which allow us to identify any given 
state in the nullcone or in the third secant variety as a point of one of the $47$ varieties described in the paper.
These $47$ varieties correspond to $47$ non-equivalent entanglement patterns, which reduce to $15$ different 
classes if we allow permutations of the qubits.
 \end{abstract}

\maketitle

\section{Introduction}

Entanglement of multi-qubit systems is a central subject in Quantum Information Theory. 
It plays an important role in applications in the field of quantum information such as quantum cryptography, quantum computation, 
quantum teleportation \cite{HHHH}. 
Recently entanglement was involved in surprising theoretical bridges like the correspondence 
between entanglement measures and string theoretic formulae for black hole entropy, leading to what is now known
as the blackhole/qubit correpondence \cite{BDD,BDDER2,BDL,Levay}. 

The question of understanding entanglement patterns of
 multipartite systems  has been investigated by various authors in the past decade
\cite{Dur,VDMV,My,chen, LLS, BDD, BDDER,HLT}.
The case of three qubits, the first nontrivial one -- denoted here as the $2\times 2\times 2$ system --, has
been solved by D\"ur  {\em et al.}\cite{Dur}  more than ten years ago, and is equivalent to the
classification of binary trilinear forms given by Le Paige\cite{LePai} in 1881. Even
if this classification is completely established,
the interpretation of entanglement for
three qubits is still under scrutiny\cite{BDDER,BDFMR,Levay1,Levay}. 
The mixed tripartite configurations $2\times 2\times n$ 
have been classified 
by Miyake {\em et al.}\cite{My,My2} and in a previous article\cite{HLT}, we have
obtained 
geometric descriptions of the $2\times 2\times n$ and $2\times 3\times 3$ quantum systems. 
In all of these classifications, we find only a finite number of nonequivalent entangled states, and they 
can be explicitly identified.

Compared to the 3-qubit case, the classification of entangled states of four qubits
 is a much more difficult problem. The Hilbert space of four qubits, 
$\mathcal{H}=\CC^2\otimes\CC^2\otimes \CC^2\otimes \CC^2$, contains 
infinitely many orbits under the action of the group $G=GL_2(\C)\times GL_2(\CC)\times GL_2(\CC)\times GL_2(\CC)$ of Stochastic Local 
Operations and Classical Communication (SLOCC). 
Therefore there is no hope to give a comprehensive classification as in the finite case. 
In terms of normal forms, a classification leads to forms depending on parameters, such as the ones of
Verstraete {\em et al.}\cite{VDMV}, corrected by Chterental and Djokovi\'c\cite{CD}. 
Another 
perspective is to describe a complete set of invariant and covariant polynomials to
separate non-equivalent orbits. This was achieved by Briand and the last two authors of this paper\cite{BLT,LT}. 
We may notice that the algebras of invariant and covariant polynomials are  quite large
(4 invariant polynomials and 170 covariant polynomials to generate both algebras), compared to the 9 normal forms given by Verstraete {\em et al.}. 
Moreover, even if the 
Verstraete {\em et al.} classification allows us to assign any 4-qubit state to one of the 9 families, it  
also implies  that states with different entanglement patterns can  belong to the same family.
The geometric study of four qubit states as $G$-invariant algebraic varieties will provide finer descriptions and make 
the connection between the normal forms and the invariant theory approaches. This is the purpose of this paper. 

The paper is organized as follow. In Section \ref{tools} we introduce the tools, 
from classical invariant theory and algebraic geometry, which
will be used all over the paper. 
We  recall 
what is known in terms of invariant and covariant polynomials and describe the method that will be used in our investigation. 
We recall some of the 
algebraic geometry techniques that we already used\cite{HLT} as well as
some recent results by Buczy\'nski and   Landsberg\cite{Lan2} which will guide us in the process
of  identifying the algebraic varieties.
In Section \ref{nulcone} we describe the set of nilpotent $4$-qubit states. 
This set, called the nullcone, is the algebraic variety defined as the zero set of all invariant polynomials.
We construct the $G$-subvarieties of the nullcone from the set of separable states and provide an algorithm 
to identify a given nilpotent $4$-qubit state as a point of one of those varieties.
In Section \ref{3sct} we describe subvarieties of the third secant variety which is the direct
generalization of the GHZ-state for four qubits. The third secant variety already contains an infinite 
number of orbits, and this will be the first example where our algorithmic method will have to be 
modified by some geometrical insights. 
This last step will allow us to explicitly describe a {\em geometric atlas} of the third secant variety 
(including the nullcone) made of $47$ non-equivalent $G$-varieties. 
Up to permutation of the qubits, this yields $15$ non-equivalent types of entanglement within 
the third secant variety.

We conclude the paper with general remarks and perspectives for further
investigation of the geometry of 4-qubit states outside of the third secant variety.
Partial results in this direction will be presented in a forthcoming paper\cite{HLT2}.

\subsection*{Notations}

Let  $\{|0\rangle,|1\rangle\}$ be a basis of $\CC^2$. A standard basis of the Hilbert space $\mathcal{H}=\CC^2\otimes\CC^2\otimes\CC^2\otimes\CC^2$ is given
by $|j_1\rangle\otimes|j_2\rangle\otimes |j_3\rangle\otimes |j_4\rangle$, with $0\leq j_i\leq 1$. 
That basis notation will be shortened in $|j_1j_2j_3j_4\rangle$ and a $4$-qubit state will be denoted by \[|\Psi\rangle=\sum_{0\leq j_1,j_2,j_3,j_4\leq 1} A_{j_1j_2j_3j_4}|j_1j_2j_3j_4\rangle\text{ with } A_{j_1j_2j_3j_4}\in \CC.\]
Nonzero scalar multiplication has no incidence on a state $|\Psi\rangle$ of the Hilbert space $\mathcal{H}$, therefore we will consider quantum states as 
points in the projective space $\PP^{15}=\PP(\mathcal{H})$. The set of separable states corresponds to tensors which can be factorized,
 i.e. $|\Psi\rangle=v_1\otimes v_2\otimes v_3\otimes v_4$ with $v_i=\alpha_i|0\rangle+\beta_i|1\rangle\in \CC^2$. The projectivization of that 
set  is an algebraic variety, called the Segre embedding of the product of four projective lines. It is the image of the following map: 
\[\begin{array}{cccc}
   \phi: & \PP(\CC^2)\times\PP(\CC^{2})\times\PP(\CC^2)\times\PP(\CC^{2}) & \to & \PP(\CC^{2}\otimes\CC^2\otimes\CC^2\otimes\CC^{2})\\
             & ([v_1],[v_2],[v_3],[v_4]) & \mapsto & [v_1\otimes v_2 \otimes v_3 \otimes v_4] 
  \end{array}\]
The Segre variety $X=\phi(\PP(\CC^{2})\times\PP(\CC^{2})\times\PP(\CC^2)\times\PP(\CC^{2}))$ will be denoted later on by
$X=\PP^{1}\times\PP^1\times\PP^1\times\PP^{1}\subset \PP(\mathcal{H})$. 
Once we work over $\PP(\mathcal{H})$ the group SLOCC will be equivalently replaced by 
$G=SL_2(\CC)\times SL_2(\CC) \times SL_2(\CC) \times SL_2(\CC)$ (with no risk of confusion $G$ will always denote 
the group SLOCC, which is the product of $GL_2(\CC)$ when we consider $\mathcal{H}$ and the product of $SL_2(\CC)$ when we consider $\PP(\mathcal{H})$).
The variety $\PP^1\times\PP^1\times\PP^1\times\PP^1$ is homogeneous  for the semi-simple Lie group $G$ and it 
corresponds to the orbit of the highest weight vector\cite{F-H,HLT} (which can be chosen to be $v=|0000\rangle$). 
More precisely the variety $X=\PP^1\times\PP^1\times\PP^1\times\PP^1=\PP(G|0000\rangle)$ is the unique homogeneous variety for the group $G$ in the sense that
for any $x,y\in X$ there exists $g\in G$ such that $y=g.x$. A variety $Y\subset \PP^{15}$ will be called a $G$-variety 
if for all $y\in Y$ and all $g\in G$ we have $g.y\in Y$.
A variety $Z$ will be called quasi-homogeneous if there exists an open dense orbit, i.e. there exists $z\in Z$ such that $Z=\PP(\overline{G.z})$.

 We work throughout with algebraic varieties over the field
$\CC$ of complex numbers. In particular we denote by $V$ a complex
vector space of dimension $N+1$ and $X^n\subset \PP(V)=\PP^{N}$ is a
complex projective nondegenerate variety (i.e. not contained in a hyperplane)
of dimension $n$. 
Given $x$ a smooth point of $X$, we denote by $T_x
X$ the intrinsic tangent space, $\tilde{T}_x X$ the embedded tangent
space\cite{Lan}, of $X$ at $x$. The notation $\hat{X}\subset V$ (resp. $\widehat{T}_x X$) will denote the cone over 
$X$ (resp. over $\tilde{T}_x X$) and $[v]\in \PP(V)$ will denote the projectivization of a vector $v\in V$. 
The dimension of the variety, $\text{dim}(X)$, is the dimension of the tangent space at a smooth point.
We say $x\in X$ is a {\em
  general point} of $X$ in the sense of the Zariski topology. 

\section{Toolbox: invariant theory and algebraic geometry}\label{tools}

\subsection{Invariant Theory}
\def\SLOCC{\mathrm{SLOCC}}
\def\Inv{\mathrm{Inv}}
\def\Cov{\mathrm{Cov}}
\def\tr{\mathrm{tr}}
\newtheorem{meth}{Method}[section]

In a more general setting a (pure) $k$-qudit system is an element of the Hilbert space $\mathcal H=V_1\otimes\cdots\otimes V_k$ with $V_i=\C^{n_i}$, equivalently it can be regarded as  a multilinear  form
\[A=\sum_{0\leq i_1\leq n_1}\cdots \sum_{0\leq i_k\leq n_k}a_{i_1,\dots,i_k}x^{(1)}_{i_1}\cdots x^{(k)}_{i_k}.\]
 Two qudit systems are equivalent if they belong in the same orbit for the group  $\SLOCC=GL_{n_1}(\mathbb C)\times\cdots\times GL_{n_k}(\mathbb C)$. The classification of multilinear  forms is an old and difficult problem treated generally by using classical (and more recently geometrical) invariant theory. The principle is the following: one describes polynomials (in the coefficient of the forms) which are invariant under the action of $\SLOCC$. Hence, if we have sufficiently many polynomials we can decide if two forms are equivalent by comparing their evaluations on these polynomials. In general, invariants are not sufficient to describe completely the orbits and we need more general polynomials, called concomittants. The set of polynomials invariants is obviously an algebra but its description, 
in terms of generators and syzygies (and even the calcultaion of its Hilbert series) is out of reach of any computer system in the general case. 
For our purposes,  we will  only deal with the $k$-qubit systems 
 (that is $n_i=2$ for each 
$1\leq k\leq i$). In the case of  multilinear binary forms, the knowledge of the covariant polynomials is sufficient. 
Let us recall briefly the main definitions:
The set of all invariants and covariants of a multilinear form of a given size are algebras 
$$\Inv:=S(\mathcal H)^{\SLOCC}\subsetneq \Cov:=[S(\mathcal H)\otimes S(V_1^*\oplus\cdots\oplus V_k^*)]^{\SLOCC}.$$ In the case of the binary forms we have $V_k=\C^2$ and the covariants are polynomials in the coefficients $\mathbf a=\{a_{i_1,\dots,i_k}: 0\leq i_1,\dots,i_k\leq 1\}$ of the form and in $k$ auxiliary binary variables $\mathbf x^{(j)}=\left(x_0^{(j)},x_1^{(j)}\right)$. Hence, it is a multigraded space $\Cov=\oplus_{d,d_1,\dots,d_k} \Cov_{d,d_1,\dots,d_k}$, where  $\Cov_{d,d_1,\dots,d_k}$ is the space of multihomogeneous polynomials of degree $d$ in $\mathbf A$ and of degree $d_i$ in $\mathbf x^{(i)}$. 
The subspace consisting in polynomials of degree $0$ in each 
binary variable $\mathbf x^{(i)}$ is the graded $\Inv=\bigoplus_d\Inv_d$ sub-algebra of $\Cov$.\\
The simplest covariant is the ground form $A$ itself,
and we will obtain all our covariants by using the Cayley Omega process. 
The Omega process is an algorithm based on a set of binary operators called transvectants. 
The transvection of two multibinary  forms $B$ and $C$ is defined by
\[
(B,C)^{i_1,\dots,i_k}:=\tr\Omega_{\mathbf x^{(1)}}^{i_1}\dots \Omega_{\mathbf x^{(k)}}^{i_k}
B(\mathbf{x'}^{(1)},\dots,\mathbf{x'}^{(1)})C(\mathbf{x''}^{(1)},\dots,\mathbf{x''}^{(1)})
\]
where $\Omega$ is the Cayley operator
\[
\Omega_x=\left|\begin{array}{cc}\frac\partial{\partial x_0'}&\frac\partial{\partial x_0''}\\
\frac\partial{\partial x_1'}&\frac\partial{\partial x_1''}\end{array}\right|,
\] 
and $\tr$ sends each $\mathbf x'$ and $\mathbf x''$ on $\mathbf x$ (erases ${}'$ and ${}''$).\\
In principle, for multilinear binary forms, one can compute a basis of $\Cov$ from the ground form $A$ by using only the operations $f\rightarrow (f,A)^{i_1,\dots,i_k}$.\\
In fact, the polynomials  here considered  are relative invariants for the action of $GL_2(\C)^k$, in the following sense: Let $F\in\Cov$ and $g_1,\dots,g_k\in GL_2(\C)$ we have $(g_1,\dots,g_k).F=(\det g_1)^{\ell_1}\cdots (\det g_k)^{\ell_k}F$, \emph{i.e.} $F$ is invariant under $GL_2(\C)^k$ up to the action of a global coefficient.\\
Let $B$ be a covariant and $$\mathbf \alpha=\sum_{0\leq i_1,\dots,i_k\leq 1}\alpha_{i_1,\dots,i_k}x^{(1)}_{i_1}\cdots x^{(k)}_{i_k}$$ be a form, the evaluation $B(\alpha)$ is the polynomial obtained by substituting 
$\alpha_{i_1,\dots,i_k}$ to $a_{i_1,\dots,i_k}$ in $B$.
We will set $B[\alpha]=1$ if $B(\alpha)\neq 0$ and $B[\alpha]=0$ otherwise.\\
Assume that we know a basis $\mathcal B$ of $\Cov$ and 
let $\mathcal B[\alpha]=\left(B[\alpha]\right)_{B\in\mathcal B}$. 
Note that if $\mathcal B[\alpha]\neq\mathcal B[\alpha']$ then $\alpha$ and $\alpha'$ do not belong in the same orbit. We define the equivalence relation $\alpha\sim\alpha'$ if and only if $\mathcal B[\alpha]=\mathcal B[\alpha']$, which partitions $\mathcal H$
into equivalence classes. 
If $\mathbb A\in\mathcal H/_\sim$, we will define $\mathcal B[\mathbb A]=\mathcal B[\alpha]$ for $\alpha\in\mathbb A$, we will denote also by $\tilde\alpha$ the class of $\alpha$. 
More precisely, we define the partial order $\preceq$ on $\mathcal H/_\sim$:  $\mathbb A\preceq\mathbb A'$ if and only if for each $B\in\mathcal B$, $B[\mathbb A']=0$ implies $B[\mathbb A]=0$. Let $\alpha\in\mathcal H$ be a form, the set $\alpha^{\preceq}=\bigcup_{\mathbb A\preceq \tilde\alpha}\mathbb A$ is the Zarisky closure of a union of orbits.  We will use the following method which is not an algorithm but rather an heuristic strategy:
\begin{meth}\label{Meth}
\begin{enumerate}
\item Compute a basis $\mathcal B$ of $\Cov$ (if it is possible otherwise compute a 
sufficiently large set $\mathcal B$ of linearly independent covariants).
\item Consider a finite set of forms $\mathcal F$. The  set $\mathcal F$  is assumed sufficiently large to contain the representatives of interesting orbits.
\item\label{m3} Compute the set $\{\alpha^\leq:\alpha\in \mathcal F\}$.
\item Find the geometric interpretation of the classes $\alpha^\leq$.
\item If the geometric investigation involves new classes then modify $\mathcal F$ and $\mathcal B$ and go back to (\ref{m3}).
\item Compute the inclusion graph. 
\end{enumerate}
\end{meth}
This method will be used as a starting point for our geometric interpretation. 
Furthermore, the set of covariants $\mathcal B$ will  provide an algorithm allowing to identify the orbit of a given form.

We have already applied this method to investigate the geometry of systems of $3$-particles\cite{HLT}. 
In each of these cases, we have found, for each orbit, a representative with coefficients in $\{0,1\}$. 
This suggests that we may start with 
$$\mathcal F\subseteq\mathcal E:=\left\{\sum_{0\leq i_1,i_2,i_3,i_4\leq 1}\alpha_{i_1,i_2,i_3,i_4}x^{(1)}_{i_1}x^{(2)}_{i_2}x^{(3)}_{i_3}x^{(4)}_{i_4}:
\alpha_{i_1,i_2,i_3,i_4}\in\{0,1\}\right\}.$$ 
For simplicity, we will denote the form $\alpha\in\mathcal F$ by the number $$\sum_{0\leq i_1,i_2,i_3,i_4\leq 1}\alpha_{i_1,i_2,i_3,i_4}2^{i_1+i_22+i_34+i_48}.$$
This set is certainly not sufficient to describe the  orbits for $4$-qubit systems,
 since there is no dense orbit and the normal forms have $4$ parameters \cite{BLT}, but we will see that if we restrict
to the nullcone or the third secant variety, our method does yield interesting results.\\

In a previous paper \cite{BLT}, we have computed a complete generating set of covariants. 
In  appendix \ref{AppCov} we propose an other generating system $\mathcal B$ which has more symmetries. 


Whilst the algebra of covariant polynomials seems very difficult to describe, the subalgebra of invariant polynomials is quite simple,
it is a free algebra on four generators:
\begin{enumerate}
\item One of degree 2: 
\[B=B_{0000}=a_{0000}a_{1111}-a_{1000}a_{0111}+a_{0100}a_{1011}+a_{1100}a_{0011}-a_{0010}a_{1101}
\]\[+a_{1010}a_{0101}-a_{0110}a_{1001}+a_{1110}a_{0001}\]
\item Two of degree 4:
\[L:=\left|\begin{array}{cccc}
a_{0000}&a_{0010}&a_{0001}&a_{0011}\\
a_{1000}&a_{1010}&a_{1001}&a_{1011}\\
a_{0100}&a_{0110}&a_{0101}&a_{0111}\\
a_{1100}&a_{1110}&a_{1101}&a_{1111}
\end{array}\right|\mbox{ and } 
M:=\left|\begin{array}{cccc}
a_{0000}&a_{0001}&a_{0100}&a_{0101}\\
a_{1000}&a_{1001}&a_{1100}&a_{1101}\\
a_{0010}&a_{0011}&a_{0110}&a_{0111}\\
a_{1010}&a_{1011}&a_{1110}&a_{1111}
\end{array}\right|
\]
\item One of degree 6: Set $b_{xy}:=\det\left(\dfrac{\partial^2 f}{\partial z_i\partial t_j}\right)$. This quadratic form is interpreted as a bilinear form on the three dimensional space $S^2(\C^2)$, so we can find a $3\times 3$ matrix $B_{xy}$ verifying
\[
b_{xy}=[x_0^2,x_0x_1,x_1^2]B_{xy}\left[\begin{array}{c}y_0^2\\y_0y_1\\y_1^2 \end{array}\right].
\]
The generator of degree $6$ is $D_{xy}:=\det(B_{xy})$.
\end{enumerate}
Note that we can alternatively replace $L$ or $M$ by
\[N:=-L-M=\left|\begin{array}{cccc}
a_{0000}&a_{1000}&a_{0001}&a_{1001}\\
a_{0100}&a_{1100}&a_{0101}&a_{1101}\\
a_{0010}&a_{1010}&a_{0011}&a_{1011}\\
a_{0110}&a_{1110}&a_{0111}&a_{1111}
\end{array}\right|
\]
and $D_{xy}$ by $D_{xz},\dots,D_{zt}$ defined in a similar way with respect to the variables $xz,\dots,zt$.\\


There is also another invariant polynomial which have a great interest in the context of geometry: the hyperdeterminant in the sense of Gelfand-Krapranov-Zelevinsky \cite{GKZ}. Let us recall how to compute it for the case of $4$-qubits.  First consider the quartic form:
\[
R(t):=\det\left(\dfrac{\partial^2b_{xt}}{\partial x_i\partial x_j}\right)
\]
and compute the apolar $S$ of $R$ with itself and its catalecticant $T$. More precisely, setting
\[
R(t)=\sum \binom4ic_it_0^{4-i}t_1^i,
\]
we compute
\[
S:=c_0c_4-4c_1c_3+3c_2^2\mbox{ and }T:=c_0c_2c_4-c_0c_3^2+2c_1c_2c_3-c_1^2c_4-c_2^3.
\]
The hyperdeterminant is the discriminant of $R(t)$ and is given by
\[
\Delta:=S^3-27T^2.
\]
Alternatively, $\Delta$ can be constructed from the sextic forms $L_{6000}$, $L_{0600}$, $L_{0060}$ and $L_{0006}$. First choose one of the  forms, for instance
\[
L_{6000}=\sum\binom6id_ix_0^{6-i}x_1^i,
\]
and compute the degree $2$ invariant of the sextic (see \cite{Gordan}):
\[
I_2:=d_0d_6-6d_1d_5+15d_3d_4-10d^2.
\]
We remark that $\Delta$ equals $I_2$ up to a factor:
\begin{equation}\label{Delta2I_2}
{\Delta}:=\dfrac{3}{2^{19}5^2} I_2.
\end{equation}

\begin{rem}\rm
 The algebra of invariant polynomials of four qubit states has already been used\cite{VES} to refine existing 
 classifications\cite{LLS} and provide tests to distinguish certain types of entanglement. 
 However it will be clear after the case treated in Section \ref{nulcone} (the nullcone), that the algebra of 
 invariants is not sufficient in particular when we focus on states which annihilate some generators of this algebra (Sections \ref{nulcone} and \ref{3sct}).
\end{rem}

\subsection{Geometry}
The geometric interpretations which will be given in Sections \ref{nulcone} and \ref{3sct} 
are based on the construction of auxiliary varieties from the 
variety of separable states $X=\PP^1\times\PP^1\times\PP^1\times \PP^1$. 
There will be two types of constructions. The first type will consist in 
building varieties from $X$ by taking first and second order derivatives of curves of $X$. 
Those constructions are mostly  inspired 
by a recent paper of Buczy\'nski and Landsberg\cite{Lan2} where the authors 
provide a precise analysis of the normal forms of tensors which are limiting points of rank three tensors, 
i.e. points of the third secant variety (see below). 
We will use their terminology to name the new varieties. 
In the second type of construction, entangled states will be obtained by 
linear combinations of two states. This is the construction 
by joins and secants which we already explored in our previous article\cite{HLT}. 
One of the most important feature of quantum mechanics is the superposition principle, whose 
 geometric counterpart is provided precisely by the 
auxiliary varieties discussed in this section. Indeed, 
it was first noticed by Brody {\em et al.}\cite{brody1}
that for $\hat{x},\hat{y}\in \mathcal{H}$, the projective line $\PP^1_{xy}\subset \PP(\mathcal{H})$ -- the unique line in
 $\PP(\mathcal{H})$ containing $[\hat{x}]$ and $[\hat{y}]$ --   represents all possible superpositions of the states $\hat{x}$, $\hat{y}\in \mathcal{H}$. It is clear that 
the second construction by joins and secants corresponds to superposition of states.
But the first construction is also linked to the superposition principle,
because, as  will be emphasized,
 tangent lines are limits of  secant lines, i.e. states built by the first and second order information will arise as limits of superpositions.

\subsubsection{Tangential varieties, or how to build new states from first and second order information}

For $X$ a smooth projective variety, recall the definition of the tangential variety,
\[\tau(X)=\bigcup_{x\in X} \tilde{T}_x X\]

The tangential variety contains the first order information in the sense that it can be recovered by taking first derivatives of curves in $X$.  Consider a smooth curve $x(t)\subset X$ with
$x(0)=x$, then $\widehat{x}'(0)\in \widehat{T}_x X$, i.e. $x'(0) \in \tau(X)$.
On the other hand any $v\in \tau(X)$ belongs to a tangent space $\tilde{T}_x X$ and we can take 
a smooth curve $x(t)$ such that $x(0)=x$ and $x'(0)=v$. This observation leads to the following alternative definition of $\tau(X)$.
\[\tau(X)=\overline{\{x'(0), \text{ where } x(t)\subset X \text{ is a curve}\}}.\]

\begin{ex}\rm\label{deriv}
 In the context of four qubits, we have $X=\PP^1\times\PP^1\times\PP^1\times\PP^1$ and a smooth curve of $X$ 
will be  $x(t)=[e_1(t)\otimes e_2(t)\otimes e_3(t)\otimes e_4(t)]$ with $e_i(t)\in \CC^2$. Without loss of generality
we may assume, up to a change of basis, that $e_i(0)=|0\rangle$ and $e_i'(0)=|1\rangle$ (here we suppose the curve is general and
the vectors $e_i$ and $e_i'$ are not colinear for $t=0$). Let us calculate $x'(t)$. The Leibnitz rule gives
\[\widehat{x}'(t)=e_1'(t)\otimes e_2(t)\otimes e_3(t)\otimes e_4(t)+e_1(t)\otimes e_2'(t)\otimes e_3(t)\otimes e_4(t)+e_1(t)\otimes e_2(t)\otimes e_3'(t)\otimes e_4(t)\] \[
 +e_1(t)\otimes e_2(t)\otimes e_3(t)\otimes e_4'(t).
\]

That is, $\widehat{x}'(0)=|1000 \rangle+ |0100 \rangle+ |0010 \rangle+ |0001 \rangle$. The orbit $\PP(G.\widehat{x}')$ is the tangential variety
whose smooth points are W-states. 
The calculation of $x'(0)$ provides a description of the affine tangent space to $\widehat{X}$ at $\widehat{x}(0)=|0000\rangle$:
\[\hat{T}_{|0000 \rangle}X=\underbrace{\CC^2\otimes  |000 \rangle}_{V_1}+ \underbrace{|0 \rangle\otimes \CC^2\otimes  |00 \rangle}_{V_2}+ \underbrace{|00 \rangle\otimes \CC^2\otimes |0\rangle}_{V_3}+\underbrace{|000 \rangle\otimes\CC^2}_{V_4}\]

\end{ex}

Following Buczy\'nski and   Landsberg\cite{Lan2}  we can go further and consider the variety built from second order information
\[\text{Osc}(X)=\overline{\{ x'(0)+x''(0), \text{ where } x(t)\subset X \text{ is a curve}\}}\]
\begin{ex}\rm\label{deriv2}
 In example \ref{deriv}, let us differentiate $\widehat{x}'(t)$ to obtain a general point of $\text{Osc}(\PP^1\times\PP^1\times\PP^1\times\PP^1)$, 
\[\hat{x}'(0)+\hat{x}''(0)=|1000 \rangle+ |0100 \rangle+ |0010 \rangle+ |0001 \rangle+|1100 \rangle+ |1010 \rangle+ |1001 \rangle+ |0110 \rangle+|0101 \rangle+ |0011 \rangle\]
\end{ex}

The calculation of example \ref{deriv2} allows us to determine the affine second osculating space of $X$ at $|0000\rangle$

\[\widehat{T}^{(2)}_{|0000\rangle}X=\underbrace{\CC^2\otimes \CC^2\otimes |00\rangle}_{W_1}+\underbrace{\CC^2\otimes|0\rangle\otimes\CC^2\otimes |0\rangle}_{W_2} +\underbrace{\CC^2\otimes|00\rangle\otimes\CC^2}_{W_3}\]\[+
\underbrace{|0\rangle\otimes\CC^2\otimes\CC^2\otimes |0\rangle}_{W_4}+ \underbrace{|0\rangle\otimes\CC^2\otimes|0\rangle\otimes\CC^2}_{W_5}+\underbrace{|00\rangle\otimes\CC^2\otimes\CC^2}_{W_6} \]

This decomposition of the second osculating space is $P$-invariant, where $P$ is the subgroup of $G$ which stabilizes $|0000\rangle$, 
i.e. $P=\begin{pmatrix}
               * & *\\
               0 & *
              \end{pmatrix}\times\begin{pmatrix}
               * & *\\
               0 & *
              \end{pmatrix}\times\begin{pmatrix}
               * & *\\
               0 & *
              \end{pmatrix}\times\begin{pmatrix}
               * & *\\
               0 & *
              \end{pmatrix}$. Therefore we can define the following $G$-subvarieties:
\begin{enumerate}
 \item $\text{Osc}_{ij..}(X)=\overline{\{ x'(0)+x''(0), \text{ where } x(t)\subset X \text{ is a curve and } \widehat{x}''(0)\in W_i+W_j+\dots\}}$,
with the trivial inclusion $\text{Osc}_{J_1}(X)\subset \text{Osc}_{J_2}(X)\subset \text{Osc}(X)$ for $J_1\subset J_2$.
\item $\text{Osc}'(X)=\overline{\{x''(0),\text{ where }x(t)\subset X \text{ is a curve and } \widehat{x}''(0)\in \widehat{T}^{(2)} X\}}$ with the inclusion 
$\text{Osc}'(X)\subset \text{Osc}(X)$.
\end{enumerate}

In section \ref{nulcone} we will need to consider the $6$ subvarieties $\text{Osc}_{i}(X)$ and $4$ subvarieties $\text{Osc}_{ijk}(X)$.

Representatives of $\text{Osc}_J(X)$ are easily  determined. 
For instance a representative of $\text{Osc}_1(X)$ is the state 
$\underbrace{|1000 \rangle+ |0100 \rangle+ |0010 \rangle+ |0001\rangle}_{\in \widehat{T}_{|0000\rangle}}+\underbrace{|1100 \rangle}_{\in W_1}$ 
and 
a representative of $\text{Osc}_{123}(X)$ is the state $\underbrace{|1000 \rangle+ |0100 \rangle+ |0010 \rangle+ |0001 \rangle}_{\in \widehat{T}_x X}+\underbrace{|1100 \rangle+ |1010 \rangle+ |1001 \rangle}_{\in W_1+W_2+W_3}$.

In section \ref{3sct} we will meet the variety $\text{Osc}'(X)$ whose representative is, according to its definition, the state 
\[|1100 \rangle+ |1010 \rangle+ |1001 \rangle+ |0110 \rangle+|0101 \rangle+ |0011 \rangle\]
\begin{lemma}\label{osc}
 For $\sharp J\leq 3$ the variety $\text{Osc}_J(X)$ is quasihomogeneous.
\end{lemma}

\proof Let $P=\begin{pmatrix}
               * & *\\
               0 & *
              \end{pmatrix}\times\begin{pmatrix}
               * & *\\
               0 & *
              \end{pmatrix}\times\begin{pmatrix}
               * & *\\
               0 & *
              \end{pmatrix}\times\begin{pmatrix}
               * & *\\
               0 & *
              \end{pmatrix}\subset G$ be the stabilizer of $\widehat{x}=|0000\rangle$. 
              Let us show that $P$   acts transitively on the generic element of $\widehat{T}_xX+W_i+\dots+W_j$ if $\sharp J\leq 3$. 
            Let us consider the case $\sharp J=3$ and assume without loss of generality that $J=\{1,2,4\}$. This case 
            is the most generic one, as $W_1\cap W_2\cap W_4=\{|0000\rangle\}$.
            Then for any  $\widehat{x}'+\widehat{x}''\in  \widehat{T}_xX+W_1+W_2+W_4$ we have
            \[\widehat{x}'+\widehat{x}''=\lambda_1|0000\rangle+\lambda_2|1000\rangle+\lambda_3|0100\rangle+\lambda_4|0010\rangle
            +\lambda_5|0001\rangle
            +\lambda_6|1100\rangle+\lambda_7|1010\rangle+\lambda_8|0110\rangle\] i.e., it depends on $8$ parameters 
            (less than $8$ if $\sharp J<3$). 
            But $\text{dim}(P)=8$ 
and we check by direct calculation that $P$ acts transitively on the generic elements of $\widehat{T}_xX+W_1+W_2+W_4$.  Let $z=y'+y''$ 
be a general point of $\text{Osc}_J(X)$ then by homogeneity 
              of $X$ we can assume that $y'+y''\in \widehat{T}_{|0000\rangle} X+W_1+W_2+W_4$ 
              and by transitive action of $P$ any general point $z\in \text{Osc}_J(X)$ is in the $G$-orbit of 
              $|1000 \rangle+ |0100 \rangle+ |0010 \rangle+ |0001 \rangle+|1100 \rangle+ |1010 \rangle+ |0110 \rangle$.$\Box$

\begin{rem}\rm
 A consequence of Lemma \ref{osc} is that the varieties $\text{Osc}_J(X)$ are irreducible for $\sharp J\leq 3$. 
 In Section \ref{3sct}
 those varieties will be identified with more classical ones. 
\end{rem}

\begin{rem}\label{rem_dim_osc}\rm
 The varieties  $\text{Osc}_{ijk}(X)$ satifying the genericity condition $W_i\cap W_j\cap W_k=\{|0000\rangle\}$ will be of maximal 
 dimension among the  varieties  $\text{Osc}_J(X)$ with $\sharp J=3$. For instance it can be checked from the calculation of the tangent space 
 that $\text{dim}(\text{Osc}_{123}(X))<\text{dim}(\text{Osc}_{124}(X))$.
\end{rem}

Another type of variety constructed from first order information by Buczy\'nski and Landsberg\cite{Lan2} is the variety $Z(X)$. 
This variety is defined as follow
\[Z(X)=\{ x'(0)+y'(0), x(t), y(t)\subset X \text{ two curves such that } \PP^1_{x(0)y(0)}\subset X\}\]
It is proved\cite{Lan2} that $Z(X)$ is a closed variety non-necessarily irreducible.
\begin{ex}\rm
For the 4-qubit system, 4 components of $Z(X)$ will have to be considered. 
Indeed $X=\PP^1\times\PP^1\times\PP^1\times\PP^1$ and through
 any points $x\in X$, there are four lines contained in $X$. 
 Assume $\widehat{x}=|0000\rangle$ then $\PP(\CC^2\otimes |000\rangle)$, 
$\PP(|0\rangle\otimes \CC^2\otimes |00\rangle)$,
$\PP(|00\rangle \otimes \CC^2\otimes |0\rangle)$ and $\PP(|000\rangle\otimes\CC^2)$ correspond to the four lines contained in $X$ and passing through $x$.
Let us define $Z_1(X)$ as the component of $Z(X)$ obtained by $\widehat{x}(t)=e_1(t)\otimes e_2(t)\otimes e_3(t)\otimes e_4(t)$ and $\widehat{y}(t)=f_1(t)\otimes f_2(t)\otimes f_3(t)\otimes f_4(t)$ with 
$e_1(0)=|0\rangle$, $f_1(0)=|1\rangle$ and $e_2(0)=e_3(0)=e_4(0)=f_2(0)=f_3(0)=f_4(0)=|0\rangle$. 
A representative of $Z_1(X)$ is 
\[\underbrace{|0100\rangle+|0001\rangle}_{\in\widehat{T}_{|0000\rangle} X} +\underbrace{|1100\rangle+|1010\rangle}_{\in \widehat{T}_{|1000\rangle} X} \]
and similarly representatives for $Z_2(X), Z_3(X)$ and $Z_4(X)$ will be respectively
\begin{itemize}
 \item $\underbrace{|1000\rangle+|0010\rangle}_{\in\widehat{T}_{|0000\rangle} X} +\underbrace{|1100\rangle+|0101\rangle}_{\in \widehat{T}_{|0100\rangle} X}$
\item $\underbrace{|1000\rangle+|0100\rangle}_{\in\widehat{T}_{|0000\rangle} X} +\underbrace{|0110\rangle+|0011\rangle}_{\in \widehat{T}_{|0010\rangle} X}$
\item $\underbrace{|1000\rangle+|0100\rangle}_{\in\widehat{T}_{|0000\rangle} X} +\underbrace{|0101\rangle+|0011\rangle}_{\in \widehat{T}_{|0001\rangle} X}$

\end{itemize}
\end{ex}

\begin{rem}\rm
 More generally\cite{Lan2} a representative of $Z_1(X)$ will be of type 
 \[v+w=\underbrace{|1000\rangle+|0100\rangle+|0010\rangle+|0001\rangle}_{=v\in\widehat{T}_{|0000\rangle} X}+\underbrace{|1100\rangle+\alpha|1010\rangle+\beta|1001\rangle}_{w\in \widehat{T}_{|1000\rangle} X}\]
 where the parameters $\alpha$ and $\beta$ are introduced to avoid combinations between components of $v$ and $w$ (we can take any $\alpha$, $\beta$ such that $\alpha\neq \beta$ and $\alpha,\beta\neq 1$).
 One may notice in particular that $w\in W_1+W_2+W_3$. This description leads to the following observation :
 \[Z_1(X)=\text{Osc}_{123}(X).\]
 It tells us that $Z_1(X)$ will contain the varieties $\text{Osc}_1(X)$, $\text{Osc}_2(X)$ and $\text{Osc}_3(X)$ which will be
 confirmed in  Section \ref{nulcone} when we compute the adherence graph of the nullcone. According to Remark \ref{rem_dim_osc} it also tells us
 that $Z_1(X)$ will be of dimension smaller than 
 $\text{Osc}_{124}(X)$ because $W_1\cap W_2\cap W_3=\text{span}\{|0000\rangle,|1000\rangle\}$ does not
 satisfy the genericity condition. Similarly one obtains:
 \begin{itemize}
  \item $Z_2(X)=\text{Osc}_{145}(X)$,
  \item $Z_3(X)=\text{Osc}_{246}(X)$,
  \item $Z_4(X)=\text{Osc}_{356}(X)$.
 \end{itemize}
Lemma \ref{osc} implies that the varieties $Z_i(X)$ are quasihomogeneous.
\end{rem}

\subsubsection{Join and secant varieties or how to build entangled states from the superposition of two states}

The join of two varieties
$X$ and $Y$ is the (Zariski) closure of the union of the secant lines with $x\in X$ and $y\in Y$:
\[J(X,Y)=\overline{\bigcup_{x\in X, y\in Y, x\neq y} \PP^{1}_{xy}}\]

In particular if $Y=X$ the join $J(X,X)$ is called the secant variety of $X$ and will be denoted by $\sigma_2(X)$. The secant variety of $X$ is the 
closure of the set of secant lines of $X$.

\begin{rem}\rm
For $X=\PP^1\times\PP^1\times\PP^1\times \PP^1$, a generic point $x\in \sigma_2(X)$ will be a sum, $x=y+z$ of a generic pair of points of $X$. 
Under the action of $G$ we can choose this generic pair $(\widehat{y},\widehat{z})$ to be $(|0000\rangle,|1111\rangle)$. In other words the orbit $\PP(G.(|0000\rangle+|1111\rangle))$
is a dense open orbit in $\sigma_2(X)$. It is clear from the representative $\widehat{x}=|0000\rangle+|1111\rangle$ that this open set is the set of GHZ entangled states.
\end{rem}

Going further we can consider the join of $X$ and $\sigma_2(X)$ and inductively we obtain the definition of the $s$-secant variety of $X$ as the join of $X$ and $\sigma_{s-1}(X)$.
It is not difficult to check that the variety $\sigma_s(X)$  is indeed 
the closure of the union of the linear span of $s$-tuples of points of $X$:
\[\sigma_s(X)=J(X,\sigma_{s-1}(X))=\overline{\bigcup_{x_1,\dots,x_s\in X} \PP^{s-1}_{x_1\dots x_s}}\]
where $\PP^{s-1}_{x_1\dots x_s}$ is a projective space of dimension $s-1$ passing through $x_1,\dots, x_s$.

In the case of Segre products there is a definition of {\em subsecant varieties}, first introduced in our previous article\cite{HLT}:
\begin{def }\label{jpair}
 Let $Y_i\subset\PP^{n_i}$, with $1\leq i\leq m$ be  $m$ nondegenerate varieties and let us consider 
$X=Y_1\times Y_2\times\dots\times Y_m\subset \PP^{(n_1+1)(n_2+1)\dots(n_m+1)-1}$ the corresponding Segre product. 
For $J=\{j_1,\dots,j_k\}\subset \{1,\dots,m\}$, a $J$-pair of points of $X$ will be a pair $(x,y)\in X\times X$ such that 
$x=[v_1\otimes v_2\otimes\dots\otimes{v_{j_1}}\otimes v_{j_1+1}\otimes\dots\otimes{v_{j_2}}\otimes\dots\otimes{v_{j_k}}\otimes\dots\otimes v_{m}]$
  and 
$y=[w_1\otimes w_2\otimes\dots\otimes {v_{j_1}}\otimes w_{j_1+1}\otimes\dots\otimes {v_{j_2}}\otimes\dots\otimes{v_{j_k}}\otimes\dots\otimes w_{m}]$, 
i.e. the tensors $\widehat{x}$ and $\widehat{y}$ have the same components for the indices in $J$.
 
The $J$-subsecant variety of $\sigma_2(X)$ denoted by $\sigma_2(Y_1\times\dots \times\underline{Y}_{j_1}\times \dots\times \underline{Y}_{j_k}\times \dots\times Y_m)\times Y_{j_1}\times Y_{j_2}\times\dots\times Y_{j_k}$
 is the closure of the union of line
 $\PP^1_{xy}$ with $(x,y)$ a $J$-pair of  points:

\[\sigma_2(Y_1\times\dots \times\underline{Y}_{j_1}\times \dots\times \underline{Y}_{j_k}\times \dots\times Y_m)\times Y_{j_1}\times Y_{j_2}\times\dots\times Y_{j_k}
=\overline{\displaystyle\bigcup_{(x,y)\in X\times X, (x,y) J-\text{pair of points}} \PP_{xy} ^1}\] 
\end{def }

\begin{rem}\rm
 The underlined varieties in the notation of the $J$-subsecant varieties correspond to the common components for the points which define 
a  
$J$-pair. Roughly speaking those components are the ``common factor'' of $x$ and $y$ in
the decomposition of 
$z=x+y\in  \sigma_2(Y_1\times\dots \times\underline{Y}_{j_1}\times \dots\times \underline{Y}_{j_k}\times \dots\times Y_m)\times Y_{j_1}\times Y_{j_2}\times\dots\times Y_{j_k}$. 
For instance when we consider the $\{1\}$-subsecant (respectively the $\{m\}$-subsecant) 
variety we can indeed factorize the first (respectively the last) component and we have the equality $\sigma_2(\underline{Y}_1\times Y_2\times\dots \times Y_m)\times Y_1=Y_1\times \sigma_2(Y_2\times\dots\times Y_m)$.
\end{rem}

\begin{rem}\rm
 For $J=\emptyset$, the $J$-subsecant variety is $\sigma_2(X)$.
\end{rem}

\begin{rem}\rm
 As we will see, the subsecant varieties will correspond to partially entangled states. 
\end{rem}

 Suppose  $Y\subset X$. A general point of $J(Y,X)$ will be a sum of two states $z=[\widehat{x}+\widehat{y}]$ with $x\in X$ and $y\in Y$. 
 However other points may belong on $J(X,Y)$.
Let us assume $x(t)\subset X$, $y(t)\subset Y$ and $x(t),y(t)\to y_0\in Y$. 
The projective line $\PP^1_{x(t)y(t)}$ is a secant line for all $t\neq 0$ and 
the limiting line $\PP^1 _*=\lim_{t\to 0} \PP^1_{x(t)y(t)}$ is in $J(Y,X)$.
The union of the $\PP^1 _*$ is called the projective tangent star to $X$ with respect to $Y$ at the point $y_0$,
and is denoted by $T^\star _{X,Y,y_0}$. The union of the tangent stars to $X$ with respect to $Y$
is an algebraic variety, called the variety of relative tangent stars\cite{Z2} of $X$ with respect to $Y$ \[T(Y,X)=\bigcup_{y\in Y} T^\star _{X,Y,y}\]
In particular for $Y=X$ we have $T(X,X)=\tau(X)$ the tangential variety.
The expected dimension of $J(Y,X)$ is equal to $\text{dim} (X)+\text{dim}(Y)+1$ 
(${\text{dim}(X)}$ degrees of freedom to choose $x\in X$, ${\text{dim}(Y)}$ degrees of freedom to choose $y\in Y$ and $1$ degree of freedom to choose $z\in \PP^1_{xy}$).
An important consequence of the Fulton-Hansen connectedness Theorem proved by Zak\cite{Z2} insures  that if $J(X,Y)$ has the expected dimension then $T(Y,X)$ is of dimension $\text{dim}(J(Y,X))-1$ and if
$J(X,Y)$ is of dimension less than expected then $J(Y,X)=T(Y,X)$.

\begin{ex}\label{WGHZ}\rm
  For any multipartite systems $(n_1+1)\times (n_2+1)\times\dots\times (n_k+1)$ with $k\geq 3$, we have 
$\text{dim}(\sigma_2(X))=1+2\sum_{i=0} ^k n_i$, i.e. the secant variety is of the expected dimension and therefore $\tau(X)\subsetneq \sigma_2(X)$. It means
that for tripartite and higher qudit systems the GHZ and W type always exist and the states belonging to W type are on limiting lines of states of type GHZ.
Moreover the rank of tensors corresponding to the state W is greater than the rank of tensor of the GHZ state. 
This is another way to say that mutipartite
systems with $k\geq 3$ always contain exceptional states\cite{ST}.
\end{ex}

 The relation between tangential and join varieties will be illustrated in Section \ref{3sct}.

\section{The nullcone}\label{nulcone}
In this section we investigate the geometry of the nullcone. The nullcone $\widehat{\mathcal{N}}\subset \mathcal H$ 
is defined as the set of states which annihilate all invariant polynomials.
The ring of invariant polynomials being finitely generated we have
\[\widehat{\mathcal N}=\{ |\Psi\rangle\in \mathcal{H}, B(|\psi\rangle=L(|\Psi\rangle)=M(|\Psi\rangle)=D_{xy}(|\psi\rangle)=0\}\]
As usual $\mathcal{N}\subset\PP(\mathcal H)$ will denote the projectivization of the nullcone.
We apply Method \ref{Meth} by first running our algorithm and then establish the geometric interpretations.
\subsection{Computing the adherence graph}\label{nulcone_inv}
\begin{figure}[h]
\begin{tikzpicture}
 \matrix (mat) [matrix of nodes,ampersand replacement=\&,
     row sep=35pt,column sep=20pt, left delimiter={.}, right delimiter={.},
      nodes={minimum height=0.5cm,minimum width=1.5cm }]{
         \tiny \bf Genuine entanglement   \& 65511 \& 65218 \&       \& 65271 \& 65247\\
               \& 65508 \& 64762 \&       \& 65506 \& 65482\\
         64700 \& 65041 \& 65075 \&       \& 61158 \& 65109 \& 64218\\
\\
	       \&       \&       \& 59520  \&      \&        \&     \\\hline
\tiny         \bf Partial entanglement     \& 65530 \& 65518 \&       \& 65532 \& 65278\\

               \& 64160 \& 61064 \&       \& 64704 \& 59624\\
         65520 \& 65484 \& 65450 \&       \& 64764 \& 64250 \& 61166\\\hline
	\tiny\bf Unentangled states   \&       \&       \& 65535  \&      \&        \&     \\
	       \&       \&       \&    0  \&      \&        \&     \\};
\node (0) [fill=gray,opacity=0.1,rectangle,rounded corners, inner sep=0pt, fit= (mat-10-4) (mat-10-4)] {};
\node (65535) [fill=gray,opacity=0.1,rectangle,rounded corners, inner sep=0pt, fit= (mat-9-4) (mat-9-4)] {};
\node (65520) [rectangle,rounded corners, inner sep=0pt, fit= (mat-8-1) (mat-8-1)] {};
\node (65484) [rectangle,rounded corners, inner sep=0pt, fit= (mat-8-2) (mat-8-2)] {};
\node (65450) [rectangle,rounded corners, inner sep=0pt, fit= (mat-8-3) (mat-8-3)] {};
\node (64764) [rectangle,rounded corners, inner sep=0pt, fit= (mat-8-5) (mat-8-5)] {};
\node (64250) [rectangle,rounded corners, inner sep=0pt, fit= (mat-8-6) (mat-8-6)] {};
\node (61160) [rectangle,rounded corners, inner sep=0pt, fit= (mat-8-7) (mat-8-7)] {};
\node (59520)[rectangle,rounded corners, fill=gray,opacity=0.1,inner sep=0pt, fit= (mat-5-4) (mat-5-4)] {};
\node (64160) [rectangle,rounded corners, inner sep=0pt, fit= (mat-7-2) (mat-7-2)] {};
\node (61064) [rectangle,rounded corners, inner sep=0pt, fit= (mat-7-3) (mat-7-3)] {};
\node (64704) [rectangle,rounded corners, inner sep=0pt, fit= (mat-7-5) (mat-7-5)] {};
\node (59624) [rectangle,rounded corners, inner sep=0pt, fit= (mat-7-6) (mat-7-6)] {};
\node (65530) [rectangle,rounded corners, inner sep=0pt, fit= (mat-6-2) (mat-6-2)] {};
\node (65518) [rectangle,rounded corners, inner sep=0pt, fit= (mat-6-3) (mat-6-3)] {};
\node (65532) [rectangle,rounded corners, inner sep=0pt, fit= (mat-6-5) (mat-6-5)] {};
\node (65278) [rectangle,rounded corners, inner sep=0pt, fit= (mat-6-6) (mat-6-6)] {};
\node (64700) [rectangle,rounded corners, inner sep=0pt, fit= (mat-3-1) (mat-3-1)] {};
\node (65041) [rectangle,rounded corners, inner sep=0pt, fit= (mat-3-2) (mat-3-2)] {};
\node (65075) [rectangle,rounded corners, inner sep=0pt, fit= (mat-3-3) (mat-3-3)] {};
\node (61158) [rectangle,rounded corners, inner sep=0pt, fit= (mat-3-5) (mat-3-5)] {};
\node (65109) [rectangle,rounded corners, inner sep=0pt, fit= (mat-3-6) (mat-3-6)] {};
\node (64218) [rectangle,rounded corners, inner sep=0pt, fit= (mat-3-7) (mat-3-7)] {};
\node (65508) [rectangle,rounded corners, inner sep=0pt, fit= (mat-2-2) (mat-2-2)] {};
\node (64720) [rectangle,rounded corners, inner sep=0pt, fit= (mat-2-3) (mat-2-3)] {};
\node (65506) [rectangle,rounded corners, inner sep=0pt, fit= (mat-2-5) (mat-2-5)] {};
\node (65482) [rectangle,rounded corners, inner sep=0pt, fit= (mat-2-6) (mat-2-6)] {};
\node (65511) [rectangle,rounded corners, inner sep=0pt, fit= (mat-1-2) (mat-1-2)] {};
\node (65218) [rectangle,rounded corners, inner sep=0pt, fit= (mat-1-3) (mat-1-3)] {};
\node (65271) [rectangle,rounded corners, inner sep=0pt, fit= (mat-1-5) (mat-1-5)] {};
\node (65247) [rectangle,rounded corners, inner sep=0pt, fit= (mat-1-6) (mat-1-6)] {};
 \node (rect M2) [rectangle,rounded corners,fill=gray,opacity=0.1, inner sep=0pt, fit= (mat-8-1) (mat-8-7)] {};
 \node (rect M3) [rectangle,rounded corners,fill=gray,opacity=0.1, inner sep=0pt, fit= (mat-7-2) (mat-7-6)] {};
 \node (rect M5) [rectangle,rounded corners,fill=gray,opacity=0.1, inner sep=0pt, fit= (mat-6-2) (mat-6-6)] {};
 \node (rect M6) [rectangle,rounded corners,fill=gray,opacity=0.1, inner sep=0pt, fit= (mat-3-1) (mat-3-7)] {};
 \node (rect M7) [rectangle,rounded corners,fill=gray,opacity=0.1, inner sep=0pt, fit= (mat-2-2) (mat-2-6)] {};
 \node (rect M8) [rectangle,rounded corners,fill=gray,opacity=0.1, inner sep=0pt, fit= (mat-1-2) (mat-1-6)] {};
\draw (0) -- (65535);
\draw (65535)--(65520);
\draw (65535)--(65484);
\draw (65535)--(65450);
\draw (65535)--(64764);
\draw (65535)--(64250);
\draw (65535)--(61160);
\draw (65520) -- (64160);
\draw (65520) -- (64704);
\draw (65484) -- (61064);
\draw (65484) -- (64704);
\draw (65450) -- (61064);
\draw (65450) -- (64160);
\draw (64764) -- (64704);
\draw (64764) -- (59624);
\draw (61160) -- (59624);
\draw (61160) -- (61064);
\draw (64250) -- (59624);
\draw (64250) -- (64160);
\draw (64160) -- (59520);
\draw (61064) -- (59520);
\draw (64704) -- (59520);
\draw (59624) -- (59520);
\draw (64160) -- (65530);
\draw (61064) -- (65518);
\draw (64704) -- (65532);
\draw (59624) -- (65278);
\draw (65530)--(64700);
\draw (65530)--(65075);
\draw (65530)--(61158);
\draw (65518)--(64700);
\draw (65518)--(65041);
\draw (65518)--(64218);
\draw (59520)--(64700);
\draw (59520)--(65041);
\draw (59520)--(65075);
\draw (59520)--(61158);
\draw (59520)--(65109);
\draw (59520)--(64218);
\draw (65532)--(61158);
\draw (65532)--(65109);
\draw (65532)--(64218);
\draw (65278)--(65041);
\draw (65278)--(65075);
\draw (65278)--(65109);
\draw (64700)--(65508);
\draw (64700)--(64720);
\draw (65041)--(65508);
\draw (65041)--(65506);
\draw (65075)--(65508);
\draw (65075)--(65482);
\draw (61158)--(64720);
\draw (61158)--(65482);
\draw (65109)--(65506);
\draw (65109)--(65482);
\draw (64218)--(64720);
\draw (64218)--(65506);
\draw (65508)--(65511); \draw (65508)--(65218);\draw (65508)--(65271);\draw (65508)--(65247);
\draw (64720)--(65511); \draw (64720)--(65218);\draw (64720)--(65271);\draw (64720)--(65247);
\draw (65506)--(65511); \draw (65506)--(65218);\draw (65506)--(65271);\draw (65506)--(65247);
\draw (65482)--(65511); \draw (65482)--(65218);\draw (65482)--(65271);\draw (65482)--(65247);
\node [left=20pt] at (0)  {$Gr_0$};
\node [left=20pt] at (65535)  {$Gr_1$};
\node [left=20pt] at (65520)  {$Gr_2$};
\node [left=20pt] at (64160)  {$Gr_3$};
\node [left=20pt] at (59520)  {$Gr_5$};
\node [left=20pt] at (65530)  {$Gr_4$};
\node [left=20pt] at (64700)  {$Gr_6$};
\node [left=20pt] at (65508)  {$Gr_7$};
\node [left=20pt] at (65511)  {$Gr_8$};
\end{tikzpicture}
\caption{Varieties of the nullcone \label{FOrbNull}}
\end{figure}
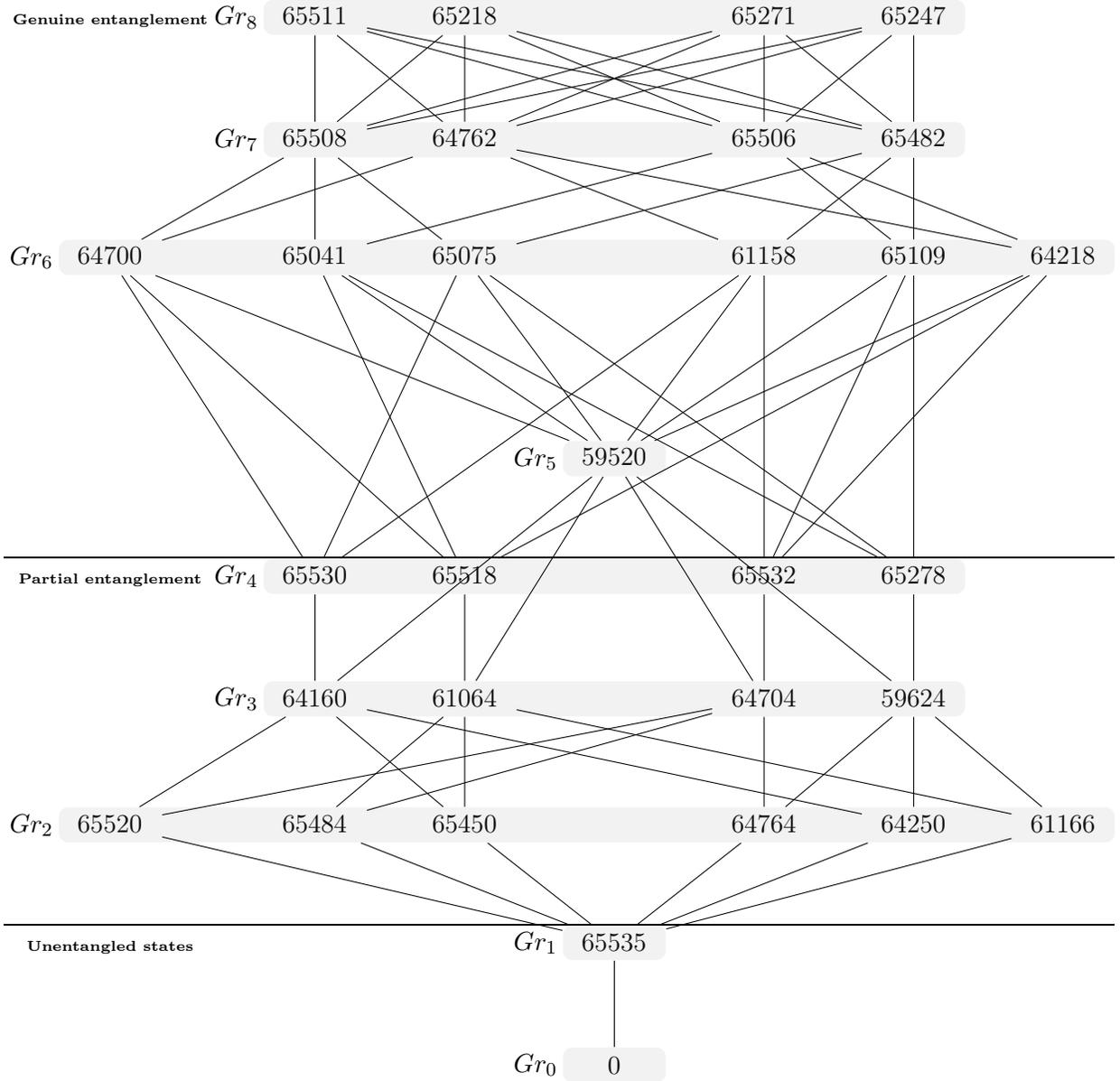

We consider the set $\mathcal F=\{\alpha\in \mathcal E: B[\alpha]=L[\alpha]=M[\alpha]=D_{xy}[\alpha]=0\}.$ Now 
the sets $\mathcal F$ and $\mathcal B$ being chosen, we let a computer algebra system find the classes, a representative for each classes and the inclusion graph.
Applied to the $11662$ forms of $\mathcal F$, we find $31$ classes whose representatives are $$\begin{array}{l}\{0,65535,         65520, 65484, 65450,        64764,  64250,  61166, 64160, 61064, 64704, 59624
 , 59520, 65530,\\ 65518, 65532, 65278, 64700, 65041, 65075, 61158, 65109, 64218,
65508, 64762, 65506, 65482,\\65511, 65218,        65271,  65247\},\end{array}$$
and obtain the inclusion graph of Fig. \ref{FOrbNull}.
In fact, we do not need all the covariants and we can summarize the results  by evaluating the following covariants on the representatives (see appendix \ref{NulCT}):
\[T:=
\begin{array}{|c|}
\hline
A\\
\hline
B_{2200},B_{2020},B_{2002},B_{0220},B_{0202},B_{0022}\\\hline
C_{3111},C_{1311} ,C_{1131}, C_{1113}\\\hline
D_{4000},D_{0400},D_{0040},D_{0004}\\\hline
D_{2200},D_{2020},D_{2002},D_{0220},D_{0202},D_{0022}\\\hline
F_{2220}^1,F_{2202}^1,F_{2022}^1,F_{0222}^1\\\hline
L_{6000},L_{0600},L_{0060},L_{0006}\\\hline
\end{array}
\]
Fig.  \ref{FOrbNull} is directly deduced from appendix \ref{NulCT}. It is interesting to remark that in this case, 
we obtain the classification without the help of the geometry (we will see in subsection \ref{GeoNull} why the 
classification is complete). Straightforwardly, the results contained in appendix \ref{NulCT} provide an algorithm to
decide to which orbit a given form belongs. \\
\begin{algo}\label{AlgoNulCone} Compute the orbit of a nilpotent form $|\Psi\rangle$.\\
{\tt Input:} A state $|\Psi\rangle$\\
{\tt Output:} The name of the orbit (according to Fig. \ref{FOrbNull}) or {\tt FAIL} if $|\Psi\rangle$ is not nilpotent.\\
{\tt If } $|\Psi\rangle$ is nilpotent {\tt then}\\
 {\color{white}......} compute $T\left[|\Psi\rangle\right]$ and compare it to Appendix B\\
{\tt else} {\tt FAIL}
\end{algo}





Observing Fig. \ref{FOrbNull}, we can define $9$ stratas of representatives whose elements have symmetric behaviors \emph{w.r.t.} their evaluations:\\
$Gr_0:=\{0\}$,\\
$Gr_1:=\{65535\}$,\\
$Gr_2:=\{65520,65484,65450,64764,64250,61166\}$,\\
$Gr_3:=\{64160,61064,64704,59624\}$,\\
$Gr_4:=\{65530,65518,65532,65278\}$,\\
$Gr_5:=\{59520\}$,\\
$Gr_6:=\{64700,65041,65075,61158,65109,64218\}$,\\
$Gr_7:=\{65508,64762,65506,65482\}$,\\
$Gr_8:=\{65511,65218,65271,65247\}$.\\
Consider the following polynomials:\\
$P_B:=B_{2200}+B_{2020}+B_{2002}+B_{0220}+B_{0202}+B_{0022},$\\
$P_C^1:=C_{3111}+C_{1311}+C_{1131}+C_{1113}$,\\
$P_C^2:=C_{3111}C_{1311}C_{1131}C_{1113}$,\\
$P_D^1:=D_{4000}+D_{0400}+D_{0040}+D_{0004}$,\\
$P_D^2:=D_{2200}+D_{2020}+D_{2002}+D_{0220}+D_{0202}+D_{0022}$,\\
$P_F:=F_{2220}^1+F_{2202}^1+F_{2022}^1+F_{0222}^1$,\\
$P_L:=L_{6000}+L_{0600}+L_{0060}+L_{0006}$.\\
We can decide to which strata a given form belongs by evaluating the vector:
\[
V:=[A,P_B,P_C^1,P_C^2,P_D^1,P_D^2,P_F,P_L].
\]
The results are summarized in Table \ref{TNullStra}, where $1$ means that the covariant does not vanish.
\begin{table}[h]
\[
\begin{array}{|c|c|}
\hline
Gr_0&[0, 0, 0, 0, 0, 0, 0, 0]\\
Gr_1&[1, 0, 0, 0, 0, 0, 0, 0]\\
Gr_2&[1, 1, 0, 0, 0, 0, 0, 0]\\
Gr_3&[1, 1, 1, 0, 0, 0, 0, 0]\\
Gr_4&[1, 1, 1, 0, 1, 0, 0, 0]\\
Gr_5&[1, 1, 1, 1, 0, 0, 0, 0]\\
Gr_6&[1, 1, 1, 1, 1, 1, 0, 0]\\
Gr_7&[1, 1, 1, 1, 1, 1, 1, 0]\\
Gr_8&[1, 1, 1, 1, 1, 1, 1, 1]
\\\hline
\end{array}
\]
\caption{The values of $V[\alpha]$ on each strata of the nullcone \label{TNullStra}}
\end{table}

\subsection{The geometry of the nullcone\label{GeoNull}}

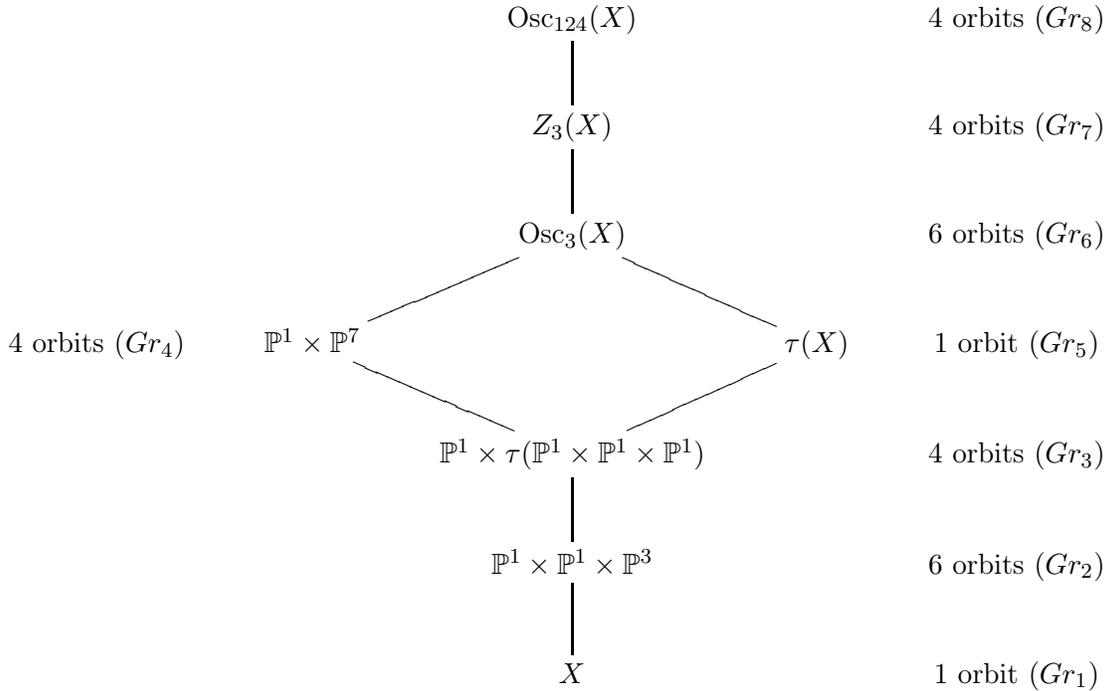
\begin{figure}[!h]
\[\xymatrix{& & \text{Osc}_{124}(X)\incl[d]&& 4 \text{ orbits } (Gr_8) \\
          && Z_3(X)\incl[d] & &4 \text{ orbits } (Gr_7) \\
           &       &      \text{Osc}_{3}(X)\incl[dr]\incl[dl]&  &6 \text{ orbits } (Gr_6)\\
   4 \text{ orbits } (Gr_4) &    \PP^1\times \PP^7 \incl[dr]           &      & \tau(X)\incl[dl]&   1 \text{ orbit } (Gr_5)  \\
            &      & \PP^1\times \tau(\PP^1\times\PP^1\times \PP^1)\incl[d]& &4 \text{ orbits } (Gr_3) \\
         & &  \PP^1\times \PP^1\times \PP^3   & &6 \text{ orbits }   (Gr_2)\\
          &        & X \incl[u] & &1 \text{ orbit }  (Gr_1)
}\]
\caption{Sketch of the stratification of the nullcone by $G$-algebraic varieties. Only one variety is given 
for each group $Gr_i$,  other representatives of each group correspond to permutations.}\label{figure_nul_cone}
\end{figure}

We now translate   the previous calculations in a geometric description of the $G$-orbits of the projectivized nullcone $\mathcal{N}\subset \PP^{15}$.
\begin{theorem}\label{thnulcone}
 The nullcone $\mathcal{N}^{11}\subset \PP^{15}$ is the union of 4 irreducible algebraic varieties of dimension $11$ and contains $30$ non-equivalent classes of (entangled) states\footnote{The set of separable states as well as sets of partially entangled states are part of the orbits.} .
All of those classes are algebraic varieties which can be built up by geometric constructions from the set of 
separable states $X=\PP^1\times\PP^1\times\PP^1\times\PP^1\subset \PP^{15}$.
The identifications of those algebraic varieties are given in Table \ref{table_nul_cone1} (genuine entangled states) and 
Table \ref{table_nul_cone} (partially entangled states).
The stratification of the nullcone by those $G$-varieties is  the adherence graph of Figure \ref{FOrbNull} (without the trivial orbit)
 and it is sketched in geometric terms in Figure \ref{figure_nul_cone}. 
\end{theorem}

\proof The action of $G=SL_2\times SL_2\times SL_2\times SL_2$ on $\PP(\mathcal{H})$ is not finite but 
it is known from Kac's work\cite{Kac} that $G$ 
acts with finitely many orbits on $\mathcal{N}$. More precisely according to Kostant-Sekiguchi Theorem\cite{DLS} 
there are 30 $G$-orbits in $\mathcal{N}$ ($31$ in $\widehat{\mathcal{N}}$ when adding the trivial orbit). 
Therefore the orbits identified by the calculation of Section \ref{nulcone_inv} exhaust all orbits of the nullcone.
To identify the orbit closures calculated in  Section \ref{nulcone_inv} with the varieties of Tables \ref{table_nul_cone1} and \ref{table_nul_cone} 
we first
check with the algorithm that the representative of a variety is a point of the corresponding orbit and thus the $G$-orbit of the representative is contained in the given orbit.
But all varieties involved in Tables \ref{table_nul_cone1} and \ref{table_nul_cone} are quasihomogneous (it is proved for $\text{Osc}_{ijk}(X)$ in Section \ref{tools} and it is
well known for the other orbits\cite{HLT}). Then comparing dimensions proves the equality between the varieties and the orbit closures. $\Box$

\begin{table}[!h]
\begin{center}
\begin{tabular}{|c|c|c|c|}
\hline
Name & Variety (orbit closure) & Normal form & dim \\
\hline
$65247$ & $\text{Osc}_{135}(X)$&$|0001\rangle+|0010\rangle+|0100\rangle+|1000\rangle+|1100\rangle+|1001\rangle+|0101\rangle$ & $11$\\
$65271$ & $\text{Osc}_{236}(X)$&$|0001\rangle+|0010\rangle+|0100\rangle+|1000\rangle+|1010\rangle+|1001\rangle+|0011\rangle$ &$11$ \\
$65218$ & $\text{Osc}_{456}(X)$&$|0001\rangle+|0010\rangle+|0100\rangle+|1000\rangle+|0110\rangle+|0101\rangle+|0011\rangle$ &$11$ \\
$65511$ & $\text{Osc}_{124}(X)$&$|0001\rangle+|0010\rangle+|0100\rangle+|1000\rangle+|1100\rangle+|1010\rangle+|0110\rangle$ & $11$\\
\hline
$65482$ & $Z_3(X)$ & $|1000\rangle+|0100\rangle+|0110\rangle+|0011\rangle$ & $10$\\
$65506$ & $Z_2(X)$ & $|1000\rangle+|0010\rangle+|1100\rangle+|0101\rangle$ & $10$\\
$64762$ & $Z_4(X)$ &$|1000\rangle+|0100\rangle+|0101\rangle+|0011\rangle$ & $10$\\
$65508$ & $Z_1(X)$ & $|0100\rangle+|0001\rangle+|1100\rangle+|1010\rangle$  & $10$\\
\hline
$64218$ & $\text{Osc}_{5}(X)$ & $|0000\rangle+|1000\rangle+|0100\rangle+|0010\rangle+|0001\rangle+|0101\rangle$ & $9$\\
$65109$ & $\text{Osc}_{4}(X)$ & $|0000\rangle+|1000\rangle+|0100\rangle+|0010\rangle+|0001\rangle+|0110\rangle$ & $9$\\
$61158$ & $\text{Osc}_{6}(X)$ & $|0000\rangle+|1000\rangle+|0100\rangle+|0010\rangle+|0001\rangle+|0011\rangle$ & $9$\\
$65075$ & $\text{Osc}_{2}(X)$ & $|0000\rangle+|1000\rangle+|0100\rangle+|0010\rangle+|0001\rangle+|1010\rangle$ & $9$\\
$65041$ & $\text{Osc}_{1}(X)$ & $|0000\rangle+|1000\rangle+|0100\rangle+|0010\rangle+|0001\rangle+|1100\rangle$& $9$\\
$64700$ & $\text{Osc}_{3}(X)$ & $|0000\rangle+|1000\rangle+|0100\rangle+|0010\rangle+|0001\rangle+|1001\rangle$ & $9$\\
\hline
$59520$ & $\tau(\PP^1\times\PP^1\times\PP^1\times\PP^1)$ & $|0001\rangle+|0010\rangle+|0100\rangle+|1000\rangle$ & $8$\\
\hline
\end{tabular}
\caption{Genuine entangled states ($G$-orbits) of the nullcone, their geometric identifications (varieties), their 
representatives and the dimensions of the varieties.}\label{table_nul_cone1}
\end{center}
\end{table}
\begin{table}
\begin{center}
\begin{tabular}{|c|c|c|c|}
\hline
Name & Variety (orbit closure) & Normal form & dim \\
\hline
$65278$ & $\PP^7\times\PP^1$ & $|0000\rangle+|1110\rangle$ & $8$\\
$65532$ & $\PP^1\times\PP^7$ & $|0000\rangle+|0111\rangle$ & $8$\\
$65518$ & $\sigma(\PP^1\times\PP^1\times\underline{\PP^1}\times\PP^1)\times\PP^1$&$|0000\rangle+|1101\rangle$ & $8$\\
$65530$ & $\sigma(\PP^1\times\underline{\PP^1}\times\PP^1\times\PP^1)\times\PP^1$&$|0000\rangle+|1011\rangle$ & $8$\\
\hline
$59624$ & $\tau(\PP^1\times\PP^1\times\times\PP^1)\times\PP^1$ & $|0110\rangle+|1010\rangle+|1100\rangle$ & $7$\\
$64704$ & $\PP^1\times\tau(\PP^1\times\PP^1\times\times\PP^1)$ & $|0011\rangle+|0101\rangle+|0110\rangle$ & $7$\\
$61064$ & $\tau(\PP^1\times\PP^1\times\underline{\PP^1}\times\PP^1)\times\PP^1$ & $|0101\rangle+|1001\rangle+|1100\rangle$ & $7$\\
$64160$ & $\tau(\PP^1\times\underline{\PP^1}\times\PP^1\times\PP^1)\times\PP^1$ & $|0011\rangle+|1001\rangle+|1010\rangle$ & $7$\\
\hline
$61166$ & $\sigma(\PP^1\times\PP^1)\times\PP^1\times\PP^1$ & $|0000\rangle+|1100\rangle$ & $5$\\
$64250$ & $\sigma(\PP^1\times\underline{\PP^1}\times\PP^1\times\underline{\PP^1})\times\PP^1\times\PP^1$ & $|0000\rangle+|1010\rangle$ & $5$\\
$64764$ & $\PP^1\times\sigma(\PP^1\times\PP^1)\times\PP^1$ & $|0000\rangle+|0110\rangle$ & $5$\\
$65450$ & $\sigma(\PP^1\times\underline{\PP^1}\times\underline{\PP^1}\times\PP^1)\times\PP^1\times\PP^1$ & $|0000\rangle+|1001\rangle$ & $5$\\
$65484$ & $\PP^1\times \sigma(\PP^1\times\underline{\PP^1}\times\PP^1)\times\PP^1$ & $|0000\rangle+|0101\rangle$ & $5$\\
$65520$ & $\PP^1\times\PP^1\times \sigma(\PP^1\times\PP^1)$ & $|0000\rangle+|0011\rangle$ & $5$\\
\hline
$65635$ & $\PP^1\times\PP^1\times\PP^1\times \PP^1$ &$|0000\rangle$& $4$\\
\hline
\end{tabular}
\caption{Partially entangled states ($G$-orbits) of the nullcone, their geometric identifications (varieties), 
their representatives and the dimensions of the varieties.}\label{table_nul_cone}
\end{center}
\end{table}

\section{The third secant variety}\label{3sct}
In this section we look at four qubit systems which belong to $\sigma_3(\PP^1\times\PP^1\times\PP^1\times\PP^1)$, the third secant variety 
of the set of separable states. The third secant variety of the Segre product of four projective lines is an 
algebraic variety of dimension\cite{CGG} $13$ defined by the vanishing 
of two invariant polynomials\cite{CD}:

\[\sigma_3(\PP^1\times\PP^1\times\PP^1\times\PP^1)=\{[|\Psi\rangle]\in\PP^{15}, L(|\Psi\rangle)=M(|\Psi\rangle)=0\}\]
It is also the projectivized set of tensors which are limits of tensors of rank three\cite{Lan3}.
\subsection{Computing the adherence graph}\label{comput_sec}

In this section the results obtained by our method depend on the choice of the invariant polynomials. The fact that $L$ and $M$ carry
geometrical information (they define the third secant variety) plays a particular role which will facilitate the geometric identifications. 
Looking with a computer algebra system for the classes of the set $\mathcal F=\{\alpha\in \mathcal E: L(\alpha)=M(\alpha)=0\}$ we find $17$ new orbits. 
This example illustrate the fact that our method is not really an algorithm since the interest of the results 
depend on the choice of the covariants. Here, it is better to use the invariants $L$ and $M$
instead the invariants $D^1_{0000}$ and $D^2_{0000}$ of appendix \ref{AppCov}.
The invariant of degree $6$, $F_{0000}$, will also be advantageously replaced by $D_ {xy}$.
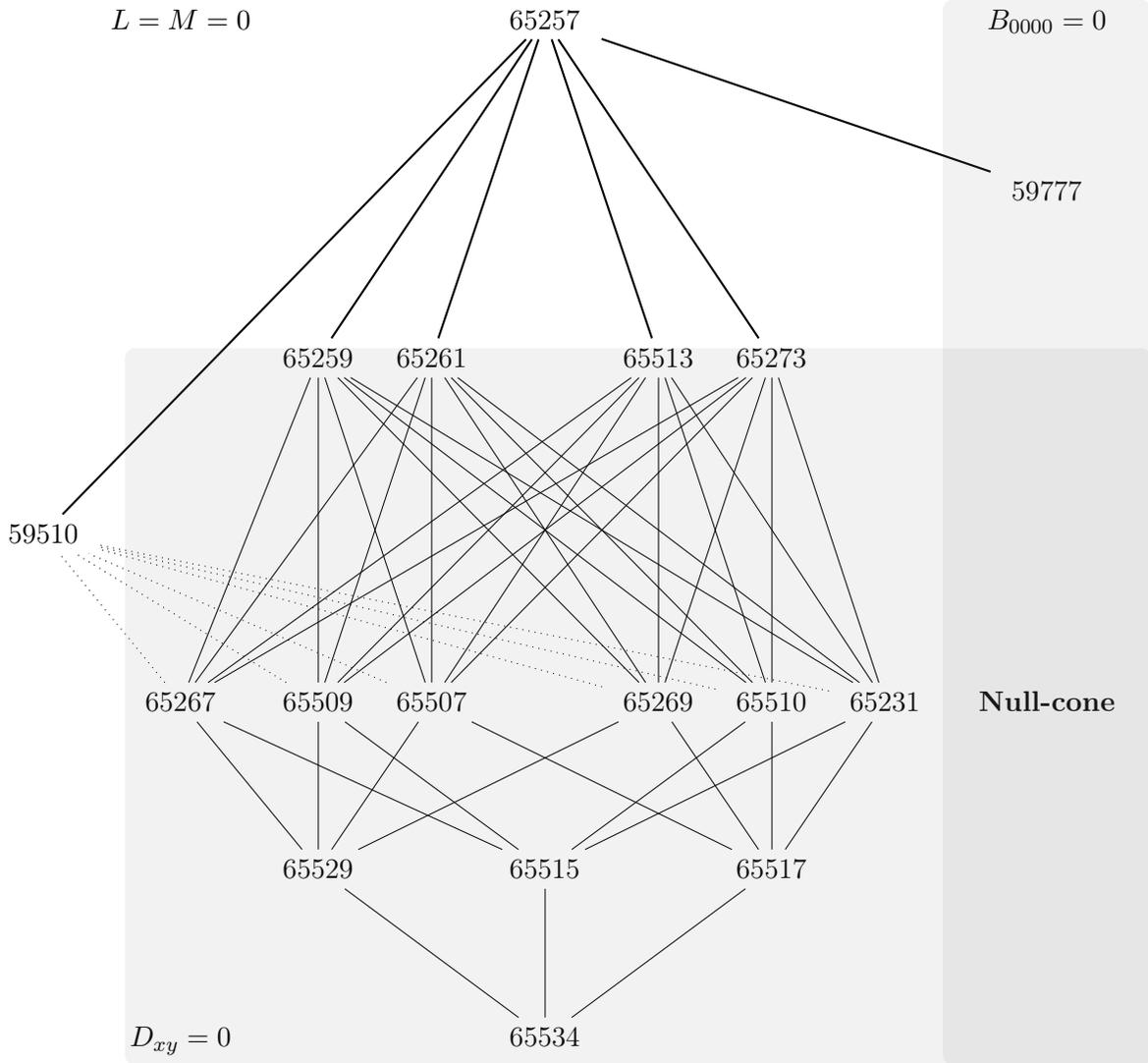
\begin{figure}[h]
\begin{center}
\begin{tikzpicture}
 \matrix (mat) [matrix of nodes,ampersand replacement=\&,
     row sep=50pt,column sep=1pt, left delimiter={.}, right delimiter={.},
      nodes={minimum height=0.5cm,minimum width=1.5cm }]{
          \&  $L=M=0$   \&       \&        \&65257 \&      \&  \& \& $B_{0000}=0$ \\
        \&      \&       \&       \&        \&      \&    \&  \& 59777\\
         \& \ \ \ \ \ \ \ \ \      \& 65259 \& 65261 \&       \& 65513 \& 65273\\
  59510\\
         \&65267 \& 65509 \& 65507 \&       \& 65269 \& 65510 \& 65231\& {\bf Null-cone}\\
          \&     \& 65529 \&       \&  65515 \&    \& 65517\\
	   \&  $D_{xy}=0$  \&       \&       \& 65534  \&      \&        \&   \& \ \ \ \ \ \ \ \ \ \ \ \ \ \ \ \ \ \ \ \ \  \\ };
\node (65257) [rectangle,rounded corners, inner sep=0pt, fit= (mat-1-5) (mat-1-5)] {};
\node (59510) [rectangle,rounded corners, inner sep=0pt, fit= (mat-4-1) (mat-4-1)] {};
\node (59777) [rectangle,rounded corners, inner sep=0pt, fit= (mat-2-9) (mat-2-9)] {};
\node (65259) [rectangle,rounded corners, inner sep=0pt, fit= (mat-3-3) (mat-3-3)] {};
\node (65261) [rectangle,rounded corners, inner sep=0pt, fit= (mat-3-4) (mat-3-4)] {};
\node (65513) [rectangle,rounded corners, inner sep=0pt, fit= (mat-3-6) (mat-3-6)] {};
\node (65273) [rectangle,rounded corners, inner sep=0pt, fit= (mat-3-7) (mat-3-7)] {};
\node (65267) [rectangle,rounded corners, inner sep=0pt, fit= (mat-5-2) (mat-5-2)] {};
\node (65509) [rectangle,rounded corners, inner sep=0pt, fit= (mat-5-3) (mat-5-3)] {};
\node (65507) [rectangle,rounded corners, inner sep=0pt, fit= (mat-5-4) (mat-5-4)] {};
\node (65269) [rectangle,rounded corners, inner sep=0pt, fit= (mat-5-6) (mat-5-6)] {};
\node (65510) [rectangle,rounded corners, inner sep=0pt, fit= (mat-5-7) (mat-5-7)] {};
\node (65231) [rectangle,rounded corners, inner sep=0pt, fit= (mat-5-8) (mat-5-8)] {};
\node (65529) [rectangle,rounded corners, inner sep=0pt, fit= (mat-6-3) (mat-6-3)] {};
\node (65515) [rectangle,rounded corners, inner sep=0pt, fit= (mat-6-5) (mat-6-5)] {};
\node (65517) [rectangle,rounded corners, inner sep=0pt, fit= (mat-6-7) (mat-6-7)] {};
\node (65534) [rectangle,rounded corners, inner sep=0pt, fit= (mat-7-5) (mat-7-5)] {};

\draw[thick,rounded corners=8pt](65257)--(59510);
\draw[thick,rounded corners=8pt](65257)--(59777);
\draw[thick,rounded corners=8pt](65257)--(65261);
\draw[thick,rounded corners=8pt](65257)--(65259);
\draw[thick,rounded corners=8pt](65257)--(65513);
\draw[thick,rounded corners=8pt](65257)--(65273);
%
%
%
%
%
%
%

\draw (65267)--(65259);\draw (65267)--(65261);\draw (65267)--(65513);\draw (65267)--(65273);
\draw (65509)--(65259);\draw (65509)--(65261);\draw (65509)--(65513);\draw (65509)--(65273);
\draw (65507)--(65259);\draw (65507)--(65261);\draw (65507)--(65513);\draw (65507)--(65273);
\draw (65269)--(65259);\draw (65269)--(65261);\draw (65269)--(65513);\draw (65269)--(65273);
\draw (65510)--(65259);\draw (65510)--(65261);\draw (65510)--(65513);\draw (65510)--(65273);
\draw (65231)--(65259);\draw (65231)--(65261);\draw (65231)--(65513);\draw (65231)--(65273);
\draw[dotted] (65231)--(59510);
\draw[dotted] (65510)--(59510);
\draw[dotted] (65269)--(59510);
\draw[dotted] (65507)--(59510);
\draw[dotted] (65509)--(59510);
\draw[dotted] (65267)--(59510);
\draw (65529)--(65267); \draw (65529)--(65509); \draw (65529)--(65507); \draw (65529)--(65269);
\draw (65515)--(65267); \draw (65515)--(65509);\draw (65515)--(65510);\draw (65515)--(65231);
\draw (65517)--(65269); \draw (65517)--(65507);\draw (65517)--(65510);\draw (65517)--(65231);
\draw (65534)--(65529);\draw (65534)--(65515);\draw (65534)--(65517);

 \node (rect M1) [rectangle,rounded corners,fill=gray,opacity=0.1, inner sep=0pt, fit= (mat-3-2) (mat-7-9)] {};
 \node (rect M2) [rectangle,rounded corners,fill=gray,opacity=0.1, inner sep=0pt, fit= (mat-1-9) (mat-7-9)] {};

\end{tikzpicture}
\end{center}
\caption{Inclusion diagram of the varieties in the third secant variety \label{OrbSec}}
\end{figure}
From these modifications, our algorithm finds $17$ new orbits:
\[\begin{array}{l}
\{65257, 59777, 59510, 65259, 65261, 65513, 65273, 65267, 65509, 65507, 65269, 65510,\\ 65231,65529, 65515, 65517, 65534 \},
\end{array}
\] 
together with the inclusion graph illustrated in Fig. \ref{OrbSec}.
\begin{rem}\label{rem5910}\rm
 The dotted edges are obtained by the calculation of the adherence graph but do not mean there are inclusions between the varieties which contains the orbits.
 Indeed the algorithm
detects a specific orbit of the third secant which do not belong to the varieties defined by $B=0$ or $D_{xy}=0$. This variety will be defined by a polynomial in $B$ and $D_{xy}$ which we should add to 
the algorithm to compute the adherence graph. In fact this missing invariant will be determined geometrically in Section \ref{3sctrev}.
\end{rem}

Obviously, we find three kinds of orbits:
\begin{enumerate}
\item $P_1:=\{65257,59510\}$ for which $B_{0000}\neq 0$ and $D_{xy}\neq 0$,
\item $P_2:=\{59777\}$ for which $B_{0000}= 0$ and $D_{xy}\neq 0$,
\item $P_3:=\{65259,\dots,65534\}$ for which $B_{0000}\neq 0$ and $D_{xy}= 0$.
\end{enumerate}
We separate the orbits of $P_1$ by testing the nullity of the components of the vector,$V'=L_{6000},\ L_{0600},\ L_{0060},\ L_{0006}\}$ (see Table \ref{V'}).
\begin{table}[h]
\[\begin{array}{|c|c|c|}\hline
\mbox{ form }& V'&\\\hline
65257&[1111]&Gr'_2\\\hline
59510&[0000]&Gr'_1\\\hline
\end{array}
\]
\caption{Values of V' \label{V'}}
\end{table} 

Now, let us show how to separate the orbits of $P_3$. First, we define the six covariants ${\bf F}_{\star\star00}:=F_{4200}+F_{2400}$, ${\bf F}_{\star0\star0}:=F_{4020}+F_{2040},\dots,\  {\bf F}_{00\star\star}:=F_{0042}+F_{0024}$. We compute the orbit of a given form in $P_3$ by evaluating the vector
\[
V''=[\mathbf F_{\star\star00},\mathbf F_{\star0\star0},\mathbf F_{\star00\star},\mathbf F_{0\star\star,0},
\mathbf F_{0\star0\star},\mathbf F_{00\star\star},L_{6000},L_{0600},L_{0060},L_{0006}].
\]
The results are summarized in Table \ref{V''} where $1$ means that the covariant does not vanish.
\begin{table}[h]
\[\begin{array}{|c|c|c|}\hline
\mbox{ form }& V''&\\\hline
65259&[1111111000]& \\
65261&[1111110100]& Gr''_4\\
65513&[1111110001]&\\
65273&[1111110010]&\\\hline
65267&[1111010000]&\\
65509&[1011110000]&\\
65507&[1110110000]&Gr''_3\\
65269&[1101110000]&\\
65510&[0111110000]&\\
65231&[1111100000]&\\\hline
65529&[1000010000]&\\
65515&[0011000000]&Gr''_2\\
65517&[0100100000]&\\\hline
65534&[0000000000]&Gr''_1\\\hline
\end{array}
\]
\caption{Evaluation of $V''[\alpha]$ \label{V''}}
\end{table} 

It is worth noticing here that two forms $\alpha$, $\alpha'$ may cancel the same invariants and take the same 
values on  the vectors $V'$ and $V''$  and still not be in the same orbit.
Recall that the third secant variety does not contain any dense orbit, it depends on two parameters. However the 
vanishing of covariants 
define  $G$-algebraic subvarieties of the third secant and two forms which cancel the same invariants and covariants will be points 
of the same $G$-algebraic variety. If the corresponding variety is quasihomogeneous then the two forms are in the same orbit.\\
Thus we have described an algorithm which can recognize points of $17+30$ subvarieties of the third secant variety.

\begin{algo}\label{AlgThirdSec} Compute the orbit of a state in the third secant\\
{\tt Input}: a state $|\Psi\rangle$\\
{\tt Output}: the name of the orbit of $|\Psi\rangle$ according to Fig. \ref{FOrbNull} and \ref{OrbSec} or {\tt FAIL} if $|\Psi\rangle$ does not belong to the third secant.\\ \\
{\tt If } $L(|\Psi\rangle)=M(|\Psi\rangle)=0$ {\tt then}\\
{\color{white} ......}{\tt If } $B_{0000}(|\Psi\rangle)=0$ {\tt then}\\
 {\color{white} ......}{\color{white} ......}{\tt If } $D_{xy}(|\Psi\rangle)=0$ {\tt then}\\
 {\color{white} ......}{\color{white} ......}{\color{white} ......} use Algo \ref{AlgoNulCone}\\
{\color{white} ......}{\color{white} ......}{\tt else} {\tt return} $59777$\\
{\color{white} ......}{\tt else}\\
 {\color{white} ......}{\color{white} ......}{\tt If} $D_{xy}(|\Psi\rangle)=0$ {\tt then}\\
{\color{white} ......}{\color{white} ......} {\color{white} ......} Evaluate $V''$ on $|\Psi\rangle$ and compare to table \ref{V''}\\
{\color{white} ......}{\color{white} ......}{\tt else} Evaluate $V'$ on $|\Psi\rangle$ and compare to table \ref{V'}\\
{\tt else} {\tt FAIL}
\end{algo}
The geometric identification of those varieties will be carried on in the next section.





\begin{figure}[h]
\begin{center}
\begin{tikzpicture}
 \matrix (mat) [matrix of nodes,ampersand replacement=\&,
     row sep=50pt,column sep=10pt, left delimiter={.}, right delimiter={.},
      nodes={minimum height=0.5cm,minimum width=1.5cm }]{
65259\& 65261\& 65513\& 65273\&\ \ \\
     \&      \&      \&       \& 59777\\ 
65511\& 65218\& 65271\& 65247\\};
\node (p65259) [rectangle,rounded corners, inner sep=0pt, fit= (mat-1-1) (mat-1-1)] {};
\node (p65261) [rectangle,rounded corners, inner sep=0pt, fit= (mat-1-2) (mat-1-2)] {};
\node (p65513) [rectangle,rounded corners, inner sep=0pt, fit= (mat-1-3) (mat-1-3)] {};
\node (p65273) [rectangle,rounded corners, inner sep=0pt, fit= (mat-1-4) (mat-1-4)] {};
\node (p59777) [rectangle,rounded corners, inner sep=0pt, fit= (mat-2-5) (mat-2-5)] {};
\node (p65511) [rectangle,rounded corners, inner sep=0pt, fit= (mat-3-1) (mat-3-1)] {};
\node (p64762) [rectangle,rounded corners, inner sep=0pt, fit= (mat-3-2) (mat-3-2)] {};
\node (p65506) [rectangle,rounded corners, inner sep=0pt, fit= (mat-3-3) (mat-3-3)] {};
\node (p65482) [rectangle,rounded corners, inner sep=0pt, fit= (mat-3-4) (mat-3-4)] {};
\draw (p65511)--(p65513);\draw (p65511)--(p59777);
\draw (p64762)--(p65259);\draw (p64762)--(p59777);
\draw (p65506)--(p65273);\draw (p65506)--(p59777);
\draw (p65482)--(p65261);\draw (p65482)--(p59777);
 \node (Gr''4) [rectangle,rounded corners,fill=gray,opacity=0.1, inner sep=0pt, fit= (mat-1-1) (mat-1-4)] {};
 \node (Gr8) [rectangle,rounded corners,fill=gray,opacity=0.1, inner sep=0pt, fit= (mat-3-1) (mat-3-4)] {};
\node [left=20pt] at (p65259)  {$Gr''_4$}; 
\node [left=20pt] at (p65511)  {$Gr_8$}; 
\end{tikzpicture}
\end{center}
\caption{$Gr''_4$, $59777$ and $Gr_8$\label{GR''4}}
\end{figure}
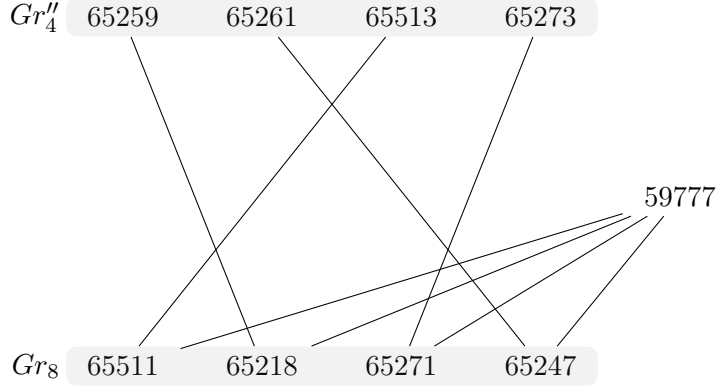
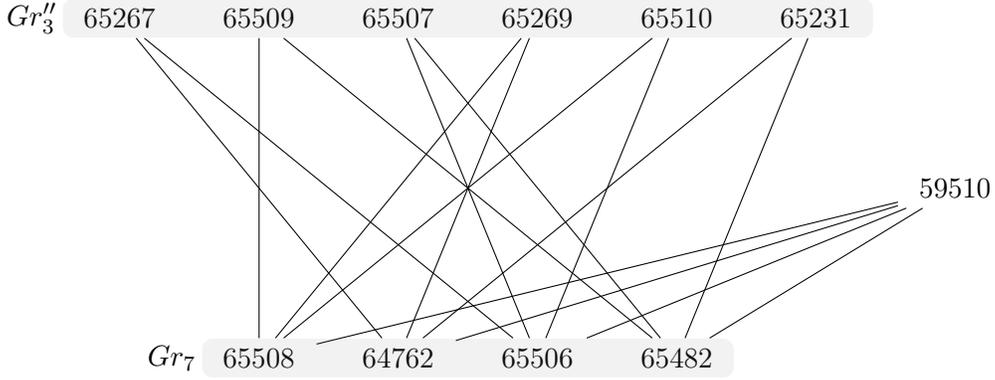
\begin{figure}[h]
\begin{center}
\begin{tikzpicture}
 \matrix (mat) [matrix of nodes,ampersand replacement=\&,
     row sep=50pt,column sep=10pt, left delimiter={.}, right delimiter={.},
      nodes={minimum height=0.5cm,minimum width=1.5cm }]{
65267\& 65509\& 65507\& 65269\& 65510\& 65231\& \\
      \&      \&      \&      \&     \&      \& 59510\\
\& 65508\& 64762\& 65506\& 65482\&\\};
\node (65267) [rectangle,rounded corners, inner sep=0pt, fit= (mat-1-1) (mat-1-1)] {};
\node (65509) [rectangle,rounded corners, inner sep=0pt, fit= (mat-1-2) (mat-1-2)] {};
\node (65507) [rectangle,rounded corners, inner sep=0pt, fit= (mat-1-3) (mat-1-3)] {};
\node (65269) [rectangle,rounded corners, inner sep=0pt, fit= (mat-1-4) (mat-1-4)] {};
\node (65510) [rectangle,rounded corners, inner sep=0pt, fit= (mat-1-5) (mat-1-5)] {};
\node (65231) [rectangle,rounded corners, inner sep=0pt, fit= (mat-1-6) (mat-1-6)] {};
\node (65508) [rectangle,rounded corners, inner sep=0pt, fit= (mat-3-2) (mat-3-2)] {};
\node (64762) [rectangle,rounded corners, inner sep=0pt, fit= (mat-3-3) (mat-3-3)] {};
\node (65506) [rectangle,rounded corners, inner sep=0pt, fit= (mat-3-4) (mat-3-4)] {};
\node (65482) [rectangle,rounded corners, inner sep=0pt, fit= (mat-3-5) (mat-3-5)] {};
\node (59510) [rectangle,rounded corners, inner sep=0pt, fit= (mat-2-7) (mat-2-7)] {};
\draw (65508)--(65509);\draw (65508)--(65269);\draw (65508)--(65510);\draw (65508)--(59510);
\draw (64762)--(65269);\draw (64762)--(65267);\draw (64762)--(65231);\draw (64762)--(59510);
\draw (65506)--(65510);\draw (65506)--(65267);\draw (65506)--(65507);\draw (65506)--(59510);
\draw (65482)--(65509);\draw (65482)--(65231);\draw (65482)--(65507);\draw (65482)--(59510);

 \node (Gr''3) [rectangle,rounded corners,fill=gray,opacity=0.1, inner sep=0pt, fit= (mat-1-1) (mat-1-6)] {};
 \node (Gr7) [rectangle,rounded corners,fill=gray,opacity=0.1, inner sep=0pt, fit= (mat-3-2) (mat-3-5)] {};
\node [left=20pt] at (65267)  {$Gr''_3$}; 
\node [left=20pt] at (65508)  {$Gr_7$}; 
\end{tikzpicture}
\end{center}
\caption{$Gr''_3$, $59510$ and $Gr_7$\label{GR''3}}
\end{figure}
\begin{figure}[h]
\begin{center}
\begin{tikzpicture}
 \matrix (mat) [matrix of nodes,ampersand replacement=\&,
     row sep=50pt,column sep=10pt, left delimiter={.}, right delimiter={.},
      nodes={minimum height=0.5cm,minimum width=1.5cm }]{
65267\& 65509\& 65507\& 65269\& 65510\& 65231 \\
      \& 65529 \&      \&  65515  \&     \&   65517 \\
64700\& 65041\& 65075\& 61158\& 65109\& 64218\\};
\node (65267) [rectangle,rounded corners, inner sep=0pt, fit= (mat-1-1) (mat-1-1)] {};
\node (65509) [rectangle,rounded corners, inner sep=0pt, fit= (mat-1-2) (mat-1-2)] {};
\node (65507) [rectangle,rounded corners, inner sep=0pt, fit= (mat-1-3) (mat-1-3)] {};
\node (65269) [rectangle,rounded corners, inner sep=0pt, fit= (mat-1-4) (mat-1-4)] {};
\node (65510) [rectangle,rounded corners, inner sep=0pt, fit= (mat-1-5) (mat-1-5)] {};
\node (65231) [rectangle,rounded corners, inner sep=0pt, fit= (mat-1-6) (mat-1-6)] {};
\node (65529) [rectangle,rounded corners, inner sep=0pt, fit= (mat-2-2) (mat-2-2)] {};
\node (65515) [rectangle,rounded corners, inner sep=0pt, fit= (mat-2-4) (mat-2-4)] {};
\node (65517) [rectangle,rounded corners, inner sep=0pt, fit= (mat-2-6) (mat-2-6)] {};
\node (64700) [rectangle,rounded corners, inner sep=0pt, fit= (mat-3-1) (mat-3-1)] {};
\node (65041) [rectangle,rounded corners, inner sep=0pt, fit= (mat-3-2) (mat-3-2)] {};
\node (65075) [rectangle,rounded corners, inner sep=0pt, fit= (mat-3-3) (mat-3-3)] {};
\node (61158) [rectangle,rounded corners, inner sep=0pt, fit= (mat-3-4) (mat-3-4)] {};
\node (65109) [rectangle,rounded corners, inner sep=0pt, fit= (mat-3-5) (mat-3-5)] {};
\node (64218) [rectangle,rounded corners, inner sep=0pt, fit= (mat-3-6) (mat-3-6)] {};

\draw (64700)--(65515);\draw (64700)--(65269);
\draw (65041)--(65529);\draw (65041)--(65510);
\draw (65075)--(65517);\draw (65075)--(65509);
\draw (61158)--(65529);\draw (61158)--(65231);
\draw (65109)--(65515);\draw (65109)--(65507);
\draw (64218)--(65517);\draw (64218)--(65267);

 \node (Gr''3) [rectangle,rounded corners,fill=gray,opacity=0.1, inner sep=0pt, fit= (mat-1-1) (mat-1-6)] {};
 \node (Gr''2) [rectangle,rounded corners,fill=gray,opacity=0.1, inner sep=0pt, fit= (mat-2-2) (mat-2-6)] {};
 \node (Gr6) [rectangle,rounded corners,fill=gray,opacity=0.1, inner sep=0pt, fit= (mat-3-1) (mat-3-6)] {};
\node [left=20pt] at (65267)  {$Gr''_3$}; 
\node [left=20pt] at (65529)  {$Gr''_2$}; 
\node [left=20pt] at (64700)  {$Gr_6$}; 
\end{tikzpicture}
\end{center}
\caption{$Gr''_3$, $Gr''_2$ and $Gr_6$\label{GR''2}}
\end{figure}
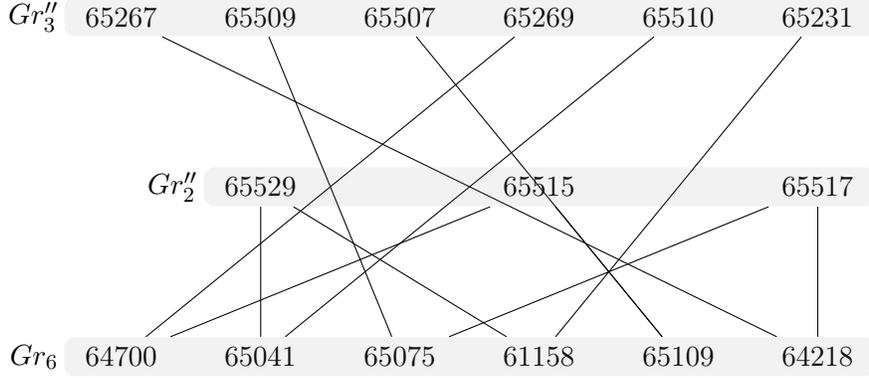
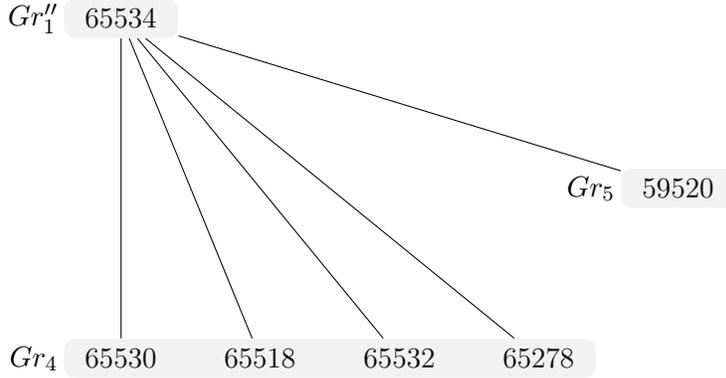
\begin{figure}[h]
\begin{center}
\begin{tikzpicture}
 \matrix (mat) [matrix of nodes,ampersand replacement=\&,
     row sep=50pt,column sep=10pt, left delimiter={.}, right delimiter={.},
      nodes={minimum height=0.5cm,minimum width=1.5cm }]{
 65534\&  \\
      \&      \&      \&       \& 59520\\
 65530\& 65518\& 65532\& 65278\&\\};
\node (65534) [rectangle,rounded corners, inner sep=0pt, fit= (mat-1-1) (mat-1-1)] {};
\node (59520) [rectangle,rounded corners, inner sep=0pt, fit= (mat-2-5) (mat-2-5)] {};
\node (65530) [rectangle,rounded corners, inner sep=0pt, fit= (mat-3-1) (mat-3-1)] {};
\node (65518) [rectangle,rounded corners, inner sep=0pt, fit= (mat-3-2) (mat-3-2)] {};
\node (65532) [rectangle,rounded corners, inner sep=0pt, fit= (mat-3-3) (mat-3-3)] {};
\node (65278) [rectangle,rounded corners, inner sep=0pt, fit= (mat-3-4) (mat-3-4)] {};

\draw (65534)--(59520);
\draw (65534)--(65278);
\draw (65534)--(65530);
\draw (65534)--(65518);
\draw (65534)--(65532);

 \node (Gr''1) [rectangle,rounded corners,fill=gray,opacity=0.1, inner sep=0pt, fit= (mat-1-1) (mat-1-1)] {};
 \node (Gr5) [rectangle,rounded corners,fill=gray,opacity=0.1, inner sep=0pt, fit= (mat-2-5) (mat-2-5)] {};
 \node (Gr4) [rectangle,rounded corners,fill=gray,opacity=0.1, inner sep=0pt, fit= (mat-3-1) (mat-3-4)] {};
\node [left=20pt] at (65534)  {$Gr''_1$}; 
\node [left=20pt] at (59520)  {$Gr_5$}; 
\node [left=20pt] at (65530)  {$Gr_4$}; 
\end{tikzpicture}
\end{center}
\caption{$Gr''_1$, $Gr_4$ and $Gr_5$\label{GR''1}}
\end{figure}

Set
\begin{eqnarray*}
\mathbf F_{42}&=&\mathbf F_{**00}+\cdots+\mathbf F_{00**}\\
\overline{\mathbf F_{**00}}&=&\mathbf F_{42}-\mathbf F_{**00}-\mathbf F_{00**}\\
\overline{\mathbf F_{*0*0}}&=&\mathbf F_{42}-\mathbf F_{*0*0}-\mathbf F_{0*0*}\\
\overline{\mathbf F_{*00*}}&=&\mathbf F_{42}-\mathbf F_{*00*}-\mathbf F_{0**0}
\end{eqnarray*}
The evaluation of the vector 
\[W=[\mathbf F_{42},\overline{\mathbf F_{**00}}\cdot\overline{\mathbf F_{*0*0}}\cdot\overline{\mathbf F_{*00*}},
\mathbf F_{**00}\mathbf F_{*0*0}\cdots\mathbf F_{00**}]
\]
allows us to determine the strata of a given form as shown by Table \ref{W}.
\begin{table}[h]\[
\begin{array}{|c|c|}\hline
\mathrm{Strata}&W\\\hline
Gr''_4& [111]\\
Gr''_3& [110]\\
Gr''_2& [100]\\
Gr''_1& [000]\\\hline
\end{array}
\]
\caption{Evaluation of $W[\alpha]$ on strata \label{W}}
\end{table}
The evaluation of the covariants allows us to describe the inclusion diagram between the stratas 
$Gr''_i$ and $Gr_j$ (see fig \ref{GR''4},\ref{GR''3},\ref{GR''2} and \ref{GR''1})

\begin{rem}\rm
 Those inclusions between groups $Gr$ and $Gr''$ will be discussed in the next section and in appendix \ref{inclusion} form the geometry perspective.
\end{rem}


\subsection{Geometric interpretations}
Let us now find out which varieties are identified by the previous calculation. 
As mentioned in Section \ref{tools},
Buczy\'nski and   Landsberg\cite{Lan2} provide a detailed and precise analysis of the normal forms of points in $\sigma_3(X)$ 
for a certain class of homogeneous varieties including the Segre products of projective spaces. They proved in particular that

\begin{theorem}(Theorem\cite{Lan2} 1.2)\label{lan}
Assume $n\geq 3$ and let $X=Seg(\PP(A_1)\times\dots\times\PP(A_n))$. Let $p=[v]\in \sigma_3(X)\diagdown\sigma(X)$.
Then $v$ has one of the following normal forms:
\begin{enumerate}
 \item $v=x+y+z$ with $[x],[y],[z]\in X$.
\item $v=x+x'+y$ with $[x],[y]\in X$ and $x'\in \widehat{T}_{[x]}X$.
\item $v=x+x'+x''$ where $[x(t)]\subset X$ is a curve and $x'=x'(0)$, $x''=x''(0)$.
\item $v=x'+y'$ where $[x]$, $[y]\in X$ are distinct points that lie on a line contained in $X$, $x'\in \widehat{T}_{[x]} X$ and $y'\in \widehat{T}_{[y]} X$.
\end{enumerate}
\end{theorem}

This theorem will serve as a guide to detect varieties among the orbits distinguished in Section \ref{comput_sec}. It says 
that
any entangled state of the third secant variety could be written as a normal form of type 1, 2, 3 or 4.
Before stating our Theorem regarding the entangled states of $\sigma_3(\PP^1\times\PP^1\times\PP^1\times\PP^1)$ 
let us make a few observations and describe some of the varieties contained in $\sigma_3(\PP^1\times\PP^1\times\PP^1\times\PP^1)$.

As already stated, in the case of $X=\PP^1\times\PP^1\times\PP^1\times\PP^1$, the third secant variety is 
of dimension $13$ and defined by $L=M=0$. 
This is an example of defective higher secant variety\cite{CGG,Lan3} (the expected dimension is $14$). 
But $\sigma_3(X)\supset J(X,\tau(X))$
 and the variety $J(X,\tau(X))$ is of dimension $13$, as it can be proved by Terracini's Lemma\cite{HLT}. 
 Thus by irreducibility
 of the varieties we have $\sigma_3(X)=J(X,\tau(X))$. It is clear\cite{Lan2} from the normal forms that point of type $2$ 
 in Theorem \ref{lan} belong to $J(X,\tau(X))$.
 In other words in the particular situation of the Segre embedding of 4 projective lines, points of type 1 and 2 are the same.
 In our Theorem other states of type 2 will appear by considering varieties $J(X,Y)$ with $Y\subset \tau(X)$ and 
 the subvarieties of $\tau(X)$ are known
 from Theorem \ref{thnulcone}.
 States of type 2 will also be  obtained by taking specific joins of varieties of 
the group $Gr_2$. An example of such a join is 
$J(\PP^3\times\PP^1\times\PP^1,\PP^1\times\PP^3\times\PP^1)$ and a general element of the cone of this variety is
$|\Psi\rangle=\underbrace{|1010\rangle+|0110\rangle}_{\in \PP^3\times\PP^1\times\PP^1}+\underbrace{|0000\rangle+|1001\rangle}_{\in\PP^1\times\PP^3\times\PP^1}$. The state $|\Psi\rangle$ 
is point of type 2 as it can be seen from the following decomposition 
\[|\Psi\rangle=\underbrace{|0000\rangle+|1010\rangle+|0110\rangle}_{\in T_{|0010\rangle} X}+\underbrace{|1001\rangle}_{\in X}\]

Other points of type 2 can be obtained by looking for forms $x'+y$ where $\widehat{x}'$ and $\widehat{y}$ are isotropic
vectors for the quadratic form befined by $B_{0000}$. Geometrically it is the same as intersecting a variety 
of points of type 2 with the projective quadric hypersurface $\QQ^{14}=\{B_{0000}=0\}\subset\PP^{15}$.
For instance we have already said that points of type 2 form a dense open subset of $\sigma_3(X)$. We can then consider \[\sigma_3^{(1)}(X)=\sigma_3(X)\cap\QQ^{14}\]
which is by construction a $G$-invariant subvariety of $\sigma_3(X)$ of codimension one in $\sigma_3(X)$.
An example of normal form for $\sigma_3^{(1)}(X)$ is \[|1111\rangle+|1000\rangle+|0100\rangle+|0010\rangle+|0001\rangle\]

Similarly one can consider $J^{(1)}(X,\tau(\PP^1\times\PP^1\times\PP^1)\times\PP^1)=J(X,\tau(\PP^1\times\PP^1\times\PP^1)\times\PP^1)\cap \QQ^{14}$
with normal form 
\[|1111\rangle+|1000\rangle+|0100\rangle+|0010\rangle\]

as well as the following varieties
\begin{center}
\begin{tabular}{|c|c|}
\hline
  $J^{(1)}(X,\tau(\PP^1\times\PP^1\times\underline{\PP^1}\times\PP^1)\times\PP^1)$ &$|1111\rangle+|1000\rangle+|0100\rangle+|0001\rangle$ \\
  \hline
  $J^{(1)}(X,\tau(\PP^1\times\underline{\PP^1}\times\PP^1\times\PP^1)\times\PP^1)$ &$|1111\rangle+|1000\rangle+|0010\rangle+|0001\rangle$  \\
  \hline
  $J^{(1)}(X,\PP^1\times\tau(\PP^1\times\PP^1\times\PP^1))$ &$|1111\rangle+|0100\rangle+|0010\rangle+|0001\rangle$  \\
  \hline
\end{tabular}
\end{center}
 
Regarding states of normal forms of type 3 we have the following lemma.
 
 \begin{lemma}\label{osc-tgt}
  Let $X=\PP^1\times\PP^1\times\PP^1\times\PP^1$ then  $\text{\em Osc}(X)=T(X,\tau(X))$.
\end{lemma}

\proof The proof is based on the same techniques as the ones developped in the paper by Buczy\'nski and   Landsberg\cite{Lan2}. 
First let us notice that $\text{Osc}(X)$ is of dimension $12$ because it is strictly contained in $\sigma_3(X)$ and it 
strictly contains the nullcone. The variety $J(X,\tau(X))$ is of dimension $13$, i.e.  the expected one. Thus 
by Zak's corollary of the Fulton-Hansen Theorem
one knows there exists an irreducible subvariety $T(X,\tau(X))$ of dimension $12$ whose 
points are on limiting secants of $J(X,\tau(X))$.
The points of $T(X,\tau(X))$ can also be seen as points on limiting $3$-planes because $J(X,\tau(X))=\sigma_3(X)$.
Let us consider three curves $x(t), y(t), z(t)$ of $X$ such that $x(0)=y(0)=z(0)=x_0\in X$ and such 
that $\PP^1_*=\lim_{t\to 0}\PP^1 _{yz}\subset \tau(X)$.
That last assumption implies that $y'(0)=z'(0)=x_1\in T_{x_0} X$. Expanding the three curves into Taylor series we get
\[\begin{array}{lll}
   x(t)& = & x_0+\dots\\
   y(t) & = & x_0 +t x_1+\dots\\
   z(t) & = & x_0+tx_1+t^2 x_2+\dots
  \end{array}\]

  with $x_2=z''(0)\in T_{x_0} ^{(2)} X$.
  
  The three plane passing through $x(t), y(t)$ and $z(t)$ can be noted as a point of $\bigwedge^3 V$
  $x(t)\wedge y(t)\wedge z(t)\in \bigwedge^3 V$ and we have
  $x(t)\wedge y(t)\wedge z(t)=x(t)\wedge(y(t)-x_0)\wedge (z(t)-x_0-tx_1)$. Thus taking the limit
  \[\lim_{t\to 0} \frac{1}{t^3} 
x(t)\wedge y(t)\wedge z(t)=x_0\wedge x_1\wedge x_2\]
The limit plane is spanned by $\tilde{x}+\tilde{x}'+\tilde{x}''$ for a curve $\tilde{x}$ such that $\tilde{x}(0)=x_0$, $\tilde{x}'(0)=x_1$ and $\tilde{x}''(0)=x_2$.
This proves $T(X,\tau(X))\subset \text{Osc}(X)$ and the equality follows by dimension argument.$\Box$

\begin{rem}\rm
 The same reasonning allows us to conclude to the following identifications
 \[\begin{array}{|l|}
 \hline
 \text{Osc}_{124}(X)=T(X,\tau(\PP^1\times\PP^1\times\times\PP^1)\times\PP^1))\\
  \text{Osc}_{456}(X)=T(X,\PP^1\times\tau(\PP^1\times\PP^1\times\times\PP^1))\\
   \text{Osc}_{135}(X)=T(X,\tau(\PP^1\times\PP^1\times\underline{\PP^1}\times\PP^1)\times\PP^1)\\
    \text{Osc}_{236}(X)=T(X,\tau(\PP^1\times\underline{\PP^1}\times\PP^1\times\PP^1)\times\PP^1)\\
    \hline
    \text{Osc}_{1}(X)=T(X,\PP^3\times\PP^1\times\PP^1) \\
    \text{Osc}_{2}(X)=T(X,\sigma(\PP^1\times\underline{\PP^1}\times\PP^1\times\underline{\PP^1})\times\PP^1\times\PP^1) \\
    \text{Osc}_{3}(X)=T(X,\sigma(\PP^1\times\underline{\PP^1}\times\underline{\PP^1}\times\PP^1)\times\PP^1\times\PP^1) \\
  \text{Osc}_{4}(X)=T(X,\PP^1\times\PP^3\times\PP^1) \\ 
   \text{Osc}_{5}(X)=T(X, \sigma(\underline{\PP^1}\times\PP^1\times\underline{\PP^1}\times\PP^1)\times\PP^1\times\PP^1) \\
    \text{Osc}_{6}(X)=T(X,\PP^1\times\PP^1\times \PP^3) \\
    \hline
 \end{array}\]
\end{rem}

Finally points of type 4 have already been discussed as they correspond to points on the varieties $Z_i(X)$.

We can now state our second Theorem.
\begin{theorem}\label{thm_3_sct}
 Algorithm \ref{AlgThirdSec} allows us to identify $17$ different classes of entangled states in $\sigma_3(X)\setminus\mathcal{N}$.
 Those classes are $G$-algebraic varieties built by superposition of states from the set of separable states.
 The states, the corresponding varieties and normal forms are given in Table \ref{table_3rd_sec2}, the inclusion among 
 the varieties
 is given by Figure \ref{OrbSec} and sketched with their
 geometric interpretations in Figure \ref{figure_3rd_sec}.
\end{theorem}

\proof We proceed similarly to our proof of Theorem \ref{thnulcone}. For each variety of Table \ref{table_3rd_sec2} we define 
normal forms from its geometric description,  as explained in Section \ref{tools}, and in 
the discussion above. Then, as a consequence of Terracini's lemma\cite{HLT}, all the varieties have the expected dimension (except $\sigma_3(X)$ as already discussed). Finally once we have the normal forms and the dimension we identify the varieties given by algorithm \ref{AlgThirdSec} with a polynomial test on the normal form. $\Box$

\begin{table}
\footnotesize 
\begin{tabular}{|c|c|c|c|}
\hline
 Name & Variety & Normal form& Dimension\\
 \hline
$6527$ & $\sigma_3(X)$& \tiny$|1111\rangle+ |0000\rangle+|1000\rangle$ 
 \tiny$+|0100\rangle+|0010\rangle+|0001\rangle$ & $13$\\
\hline
$59777$ & $\sigma_3^{(1)}(X)$ &\tiny$|1111\rangle+|1000 \rangle+ |0100 \rangle+ |0010 \rangle+ |0001 \rangle$ & $12$\\
  
       \hline
 $65513$ & $J(X,\tau(\PP^1\times\PP^1\times\PP^1)\times\PP^1)$ &\tiny$|1111\rangle+|0000\rangle+|1000\rangle+|0100\rangle+|0010\rangle$  & $12$\\
 $65261$ & $J(X,\tau(\PP^1\times\PP^1\times\underline{\PP^1}\times\PP^1)\times\PP^1)$ &\tiny$|1111\rangle+|0000\rangle+|1000\rangle+|0100\rangle+|0001\rangle$  & $12$\\
 $65273$ & $J(X,\tau(\PP^1\times\underline{\PP^1}\times\PP^1\times\PP^1)\times\PP^1)$ &\tiny$|1111\rangle+|0000\rangle+|1000\rangle+|0010\rangle+|0001\rangle$  & $12$\\
 $65259$ & $J(X,\PP^1\times\tau(\PP^1\times\PP^1\times\PP^1))$ &\tiny$|1111\rangle+|0000\rangle+|0100\rangle+|0010\rangle+|0001\rangle$  & $12$\\
 \hline
 $59510$ & $\text{Osc}'(X)$ &\tiny$|1100\rangle+|1010\rangle+|1001\rangle+|0110\rangle+|0101\rangle+|0011\rangle$ & $12$\\
 \hline
 $65507$ &  $J(\PP^3\times\PP^1\times\PP^1,\sigma(\underline{\PP^1}\times\PP^1\times\underline{\PP^1}\times\PP^1)\times\PP^1\times\PP^1)$ &\tiny $|0000\rangle+|0011\rangle+|0101\rangle+|1111\rangle$& $11$\\
$65509$ & $J(\PP^3\times\PP^1\times\PP^1,\sigma(\PP^1\times\underline{\PP^1}\times\underline{\PP^1}\times\PP^1)\times\PP^1\times\PP^1)$  &  \tiny$|0000\rangle+|1010\rangle+|0110\rangle+|1001\rangle$& $11$ \\
$65510$ &\tiny$J(\sigma(\PP^1\times\underline{\PP^1}\times\PP^1\times\underline{\PP^1})\times\PP^1\times\PP^1,   \sigma(\PP^1\times\underline{\PP^1}\times\underline{\PP^1}\times\PP^1)\times\PP^1\times\PP^1)$&\tiny$|0000\rangle+|1010\rangle+|0110\rangle+|1111\rangle$ & $11$ \\
$65231$ & $J(\PP^1\times\PP^3\times\PP^1,\sigma(\PP^1\times\underline{\PP^1}\times\PP^1\times\underline{\PP^1})\times\PP^1\times\PP^1)$  &\tiny$|0000\rangle+|0110\rangle+|0101\rangle+|1111\rangle$ & $11$ \\
$65267$ & $J(\PP^3\times\PP^1\times\PP^1,\PP^1\times\PP^3\times\PP^1)$ &\tiny$|0000\rangle+|0110\rangle+|1001\rangle+|0101\rangle$ & $11$ \\
$65269$ & $J(\PP^1\times\PP^1\times\PP^3,\sigma(\underline{\PP^1}\times\PP^1\times\underline{\PP^1}\times\PP^1)\times\PP^1\times\PP^1)$  &\tiny $|0000\rangle+|1010\rangle+|1001\rangle+|0101\rangle$ & $11$ \\
 \hline
$65529$ & $J(X,\PP^3\times\PP^1\times\PP^1)$ &\tiny $|1111\rangle+ |0000\rangle+|1000\rangle+|0100\rangle$ & $10$\\
$65517$ & $J(X,\sigma(\underline{\PP^1}\times\PP^1\times\underline{\PP^1}\times\PP^1)\times\PP^1\times\PP^1)$ &\tiny$|1111\rangle+|0000\rangle+|0100\rangle+|0001\rangle$ & $10$\\
$65515$ &  $J(X,\PP^1\times\PP^3\times\PP^1)$ &\tiny$|1111\rangle+|0000\rangle+|0100\rangle+|0010\rangle$ &$10$\\
\hline
$65534$ & $\sigma(X)$ &\tiny $|0000\rangle+|1111\rangle$ & $9$\\
\hline
\end{tabular}
\caption{Non-nilpotent entangled states of the third secant variety}\label{table_3rd_sec2}
\end{table}

\begin{rem}\rm
Those geometric interpretations allow us to get a deeper understanding of the inclusions of Figures \ref{GR''4}, \ref{GR''3}, \ref{GR''2} and \ref{GR''1}. 
For instance the inclusions of Figure \ref{GR''4} are obvious once we identify the varieties of $Gr_8$ with intersection of join varieties with $\QQ^{14}$.
The inclusions of  Figure $\ref{GR''1}$
are just the natural inclusions of the tangential and the subsecant varieties to the second secant variety (Figure \ref{GR''1bis}),
while the inclusions of Figure \ref{GR''2} between varieties of type $Gr_2''$ and $Gr_6$ are consequence of Zak's corollary
of the Fulton-Hansen Theorem (see Section \ref{tools}). More precisely the variety $J(X,\PP^3\times\PP^1\times\PP^1)$
is of the expected dimension and therefore contains a subvariety of codimension one which is $T(X,\PP^3\times\PP^1\times\PP^1)$.
But the variety $J(X,\PP^3\times\PP^1\times\PP^1)$ is identical to $J(X,\PP^1\times\PP^1\times\PP^3)$ (Figure \ref{GR''2bis}).
Finally the inclusions of Figure \ref{GR''4} correspond to inclusions of tangential varieties with their respective joins. Those tangential varieties are also 
sub-tangential varieties of $T(X,\tau(X))$ (see Figure \ref{GR''4bis}).
We gather Figures \ref{GR''4bis}, \ref{GR''2bis} and \ref{GR''1bis} with more explanations in Appendix \ref{inclusion}.
\end{rem}

\subsection{The third secant variety atlas revisited}\label{3sctrev}
It should be noticed here that the variety $\text{Osc}(X)$ does not appear in Theorem \ref{thm_3_sct}. But this variety played a crucial role
to identify subvarieties of the nullcone and subvarieties of the third secant variety.
Thus, in that example, our algorithm does not see a specific strata whose geometric feature is important to 
understand entanglement --- 
$\text{Osc}(X)$ could be interpreted as the next dimensional generalization of the W-states.

\begin{figure}[!h]
 \[\xymatrix { \sigma_3(X) \incl[d]\incl[dr]\inclu[drr]\\
    J(X,\PP^1\times\tau(\PP^1\times\PP^1\times\PP^1))\incl[d] & \sigma^{(1)} _3(X) & T(X,\tau(X))=\text{Osc}(X) \\
   J(\PP^3\times\PP^1\times\PP^1,\PP^1\times\PP^3\times\PP^1)\incl[d] & T(X,\PP^1\times\tau(\PP^1\times\PP^1\times\PP^1))\incl[u]\inclu[ur]\incl[ul] & \text{Osc}'(X)\inclu[u]\\
   J(X,\PP^1\times\PP^1\times\PP^3)\incl[d] & Z_1(X) \incl[u]\incl[ur]\incl[ul] & \\
  \sigma(X) & T(X,\PP^1\times\PP^1\times\PP^3)\incl[ul]\incl[u]& \\
         \PP^1\times \PP^7\incl[ur] \incl[u]  & \tau(X)\incl[u]\incl[ul] &  & \\
            &  \PP^1\times\tau(\PP^1\times\PP^1\times\PP^1) \incl[u]\incl[ul]       &  & \\
            &            \PP^1\times\PP^1\times\PP^3\incl[u]                                         & & \\
            &    X\incl[u]                                                 & & } \]
\caption{Extended atlas of $\sigma_3(X)$}\label{figure_3rd_sec}
\end{figure}
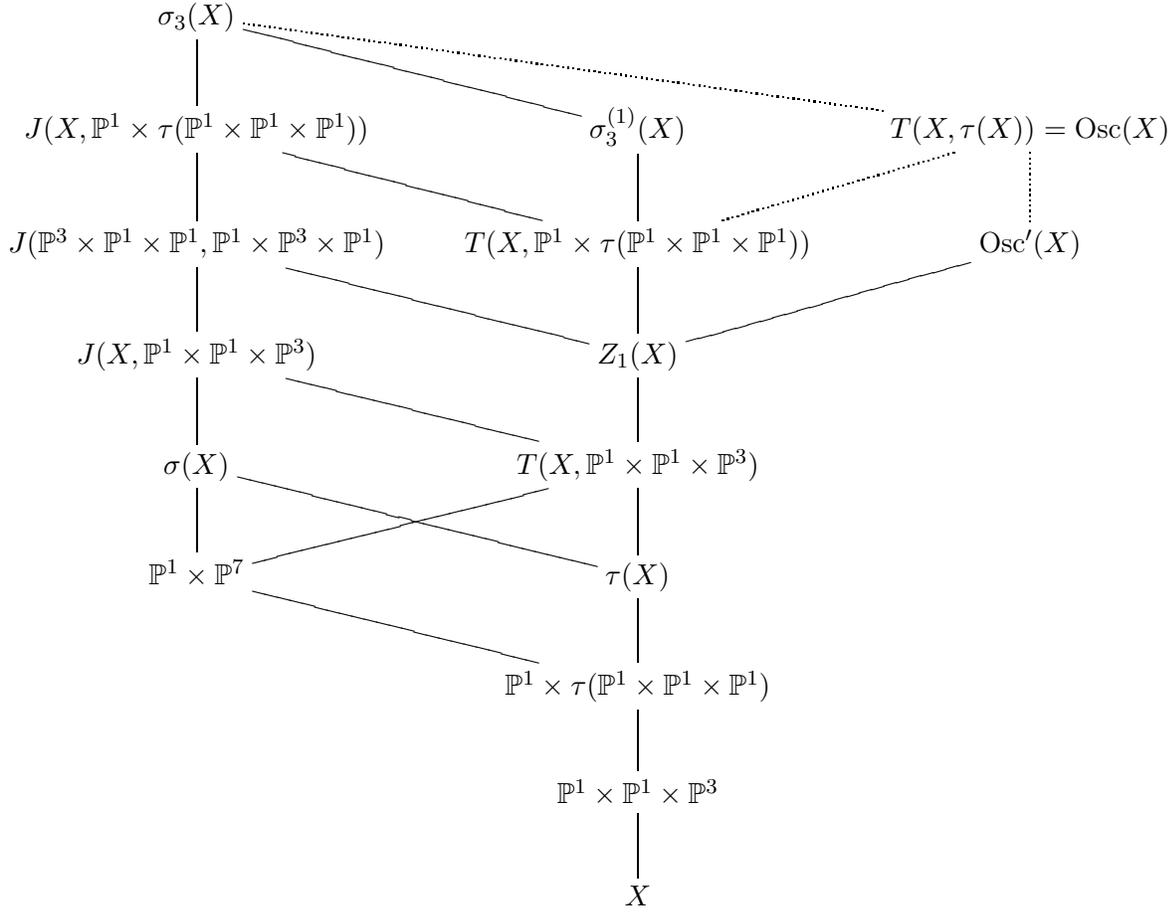

Let us sketch in Figure \ref{figure_3rd_sec} our stratification of $\sigma_3(X)$ by algebraic varieties. 
We mark by dotted edges the inclusion of the
variety $\text{Osc}(X)$ and 
we give only one variety of each group $Gr_i$ and $Gr_i''$ (the others being obtained by permutations).

Back to our description of our Method \ref{Meth}, geometry suggests here 
to introduce a new invariant to identify $\text{Osc}(X)$.
To obtain by calculation the extended
decomposition of the third secant variety one needs to introduce the new invariant \[Z=D_{xy}-\dfrac{1}{27}B^3\]
and modify our original algorithm.

\begin{figure}[h]
\begin{center}
\begin{tikzpicture}
 \matrix (mat) [matrix of nodes,ampersand replacement=\&,
     row sep=50pt,column sep=10pt, left delimiter={.}, right delimiter={.},
      nodes={minimum height=0.5cm,minimum width=1.5cm }]{  \&     \&       \&       \& {\color{gray} 65257} \&      \&  \& \&  \\
    \bf 6014  \&  \  $\Delta=0$    \&      \color{gray} 65259    \&\color{gray}  65261  \  \& \& \color{gray}65513      \& \color{gray}65273    \&  \&\color{gray} 59777\\
     \color{gray} 59510     \&      \&  \& \&       \&  \& \& \& \\
    \&     \&       \&        \& \&      \&  \& \&  \\
  $Z=0$ \&\color{gray}65267 \& \color{gray}65509 \& \color{gray}65507 \&       \& \color{gray}65269 \& \color{gray}65510 \& \color{gray}65231\& \\
};
\node (6014) [rectangle,rounded corners, inner sep=0pt, fit= (mat-2-1) (mat-2-1)] {};
\node (65257) [rectangle,rounded corners, inner sep=0pt, fit= (mat-1-5) (mat-1-5)] {};
\node (59510) [rectangle,rounded corners, inner sep=0pt, fit= (mat-3-1) (mat-3-1)] {};
\node (59777) [rectangle,rounded corners, inner sep=0pt, fit= (mat-2-9) (mat-2-9)] {};
\node (65259) [rectangle,rounded corners, inner sep=0pt, fit= (mat-2-3) (mat-2-3)] {};
\node (65261) [rectangle,rounded corners, inner sep=0pt, fit= (mat-2-4) (mat-2-4)] {};
\node (65513) [rectangle,rounded corners, inner sep=0pt, fit= (mat-2-6) (mat-2-6)] {};
\node (65273) [rectangle,rounded corners, inner sep=0pt, fit= (mat-2-7) (mat-2-7)] {};
\node (65267) [rectangle,rounded corners, inner sep=0pt, fit= (mat-5-2) (mat-5-2)] {};
\node (65509) [rectangle,rounded corners, inner sep=0pt, fit= (mat-5-3) (mat-5-3)] {};
\node (65507) [rectangle,rounded corners, inner sep=0pt, fit= (mat-5-4) (mat-5-4)] {};
\node (65269) [rectangle,rounded corners, inner sep=0pt, fit= (mat-5-6) (mat-5-6)] {};
\node (65510) [rectangle,rounded corners, inner sep=0pt, fit= (mat-5-7) (mat-5-7)] {};
\node (65231) [rectangle,rounded corners, inner sep=0pt, fit= (mat-5-8) (mat-5-8)] {};
%
%

\draw[dotted](65257)--(59777);
\draw[thick,rounded corners=8pt](65257)--(6014);
\draw[dotted](65257)--(65261);
\draw[dotted](65257)--(65259);
\draw[dotted](65257)--(65513);
\draw[dotted](65257)--(65273);

\draw[dotted] (65267)--(65259);\draw[dotted] (65267)--(65261);
\draw[dotted] (65267)--(65513);\draw[dotted] (65267)--(65273);
\draw[dotted] (65509)--(65259);\draw[dotted] (65509)--(65261);\draw[dotted] (65509)--(65513);
\draw[dotted] (65509)--(65273);
\draw[dotted] (65507)--(65259);\draw[dotted] (65507)--(65261);\draw[dotted] (65507)--(65513);
\draw[dotted] (65507)--(65273);
\draw[dotted] (65269)--(65259);\draw[dotted] (65269)--(65261);\draw[dotted] (65269)--(65513);
\draw[dotted] (65269)--(65273);
\draw[dotted] (65510)--(65259);\draw[dotted] (65510)--(65261);\draw[dotted] (65510)--(65513);
\draw[dotted] (65510)--(65273);
\draw[dotted] (65231)--(65259);\draw[dotted] (65231)--(65261);\draw[dotted] (65231)--(65513);
\draw[dotted] (65231)--(65273);
\draw[thick,rounded corners=8pt](6014)--(59510);%

%
%
%
%
%
%
\node (Gr4) [rectangle,rounded corners,fill=gray,opacity=0.1, inner sep=0pt, fit= (mat-2-1) (mat-5-1)] {};
\node (Gr5) [rectangle,rounded corners,fill=gray,opacity=0.1, inner sep=0pt, fit= (mat-2-1) (mat-5-8)] {};

\end{tikzpicture}
\end{center}
\caption{New inclusion diagram of the varieties in the third secant variety \label{OrbSec2}}
\end{figure}
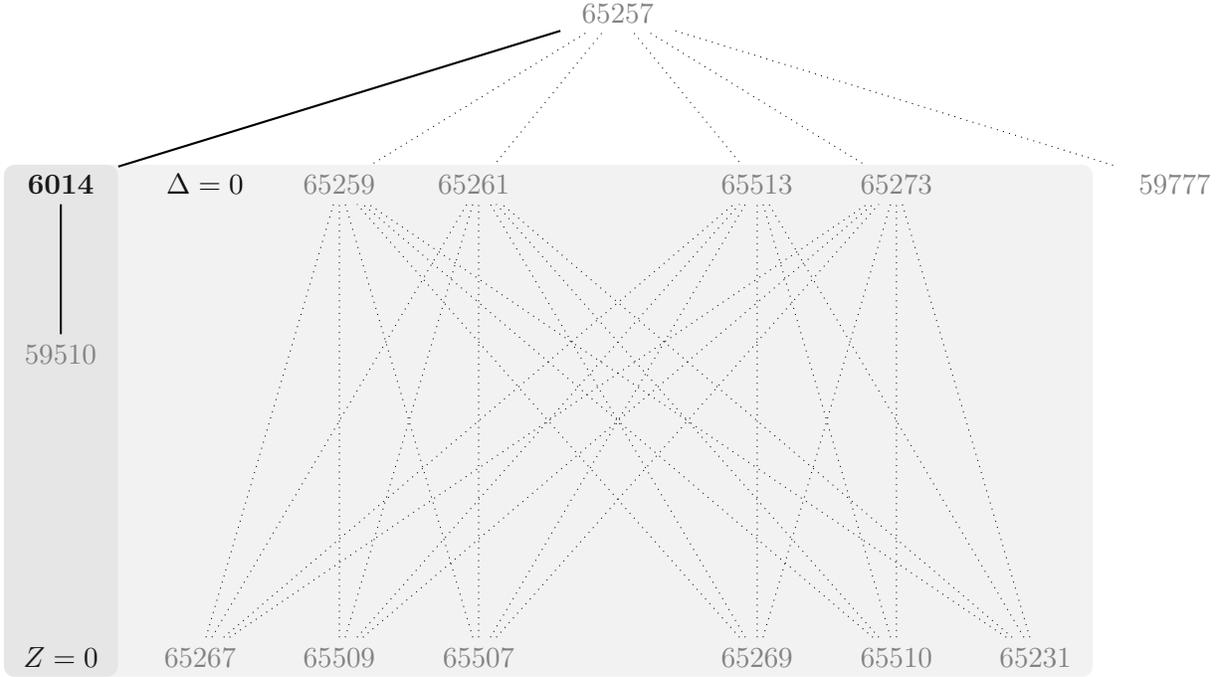

We can also use the polynomial $\Delta$ (the hyperdeterminant, see Section \ref{tools}) to slightly refine the inclusion graph. We remark 
first that for non-nilpotent forms verifying $L=M=0$, we have $D_{xy}=0$ implies
$\Delta=0$, $B_{0000}=0$ implies $\Delta\neq 0$ and from eq (\ref{Delta2I_2}) 
$L_{6000}=0$ implies $\Delta=0$. So the condition $\Delta=0$ allows only to find
one more subvariety of $65257$ whose  representative is $6014$. When $L=M=0$ and $D_{xy}\neq 0$ 
the polynomials $\Delta$ and $Z$ play the same role because $\Delta_{|L=M=0}=6912D_{xy}Z$. We will explore this type of description of $\Delta$ 
in a forthcoming paper\cite{HLT2}. As a consequence, we obtain a new atlas, and a new algorithm with $Z$, for entanglement 
types within the third secant variety (see Fig. \ref{OrbSec2} which is the computational analogue of Fig. \ref{figure_3rd_sec}) including a new
variety whose  representative is $6014$. Note also that the variety 
represented by $59510$ has only nilpotent strict subvarieties in the graph as suspected in Remark \ref{rem5910}.  

It is quite a good surprise to find out here that the missing invariant to detect $\text{Osc}(X)$ was a component of the restriction to $\sigma_3(X)$ of 
the hyperdeterminant $\Delta$. The  hypersurface given by $\Delta=0$ which
is the dual variety of $X=\PP^1\times\PP^1\times\PP^1\times\PP^1$ should play a central role to understand 
the entanglement of four qubits as already pointed out by Miyake\cite{My}.

%



\section{Conclusion}
We have investigated the geometry of  four qubit systems using the knowledge of the algebra of 
covariant polynomials and the algebraic geometry of auxiliary varieties.
Imposing conditions on the generators of the algebra of invariant polynomials, we have been  able to describe precisely
the so-called nullcone variety and the third secant variety. 
Our descriptions are geometric --- each entanglement pattern appears as an open subset of an algebraic variety,
and algorithmic --- each strata is identified by the vanishing of specific covariants.

Our motivations for studying the third secant variety of the four qubit systems had been explained
 in our previous paper\cite{HLT}. The third 
secant variety is the next higher dimensional generalization of the GHZ-state. 
It is also a case of higher secant variety investigated by geometers\cite{Lan2,CGG}.

The method (Method \ref{Meth}) applied in this paper gives a complete description of the nullcone as well as specific 
stratas of the third secant variety given by 
the algorithm and completed by geometric analysis. 
This geometric atlas of the third variety is made of $18$ ($17$ found by the algorithm and $1$ by geometric 
considerations, corresponding to the intersection of the third secant with the zero locus of the hyperdeterminant) 
classes of non-nilpotent entangled states and $29$ nilpotent entangled states
(the orbits of the nullcone without the set of separable states).
The new classes of entanglement are all built by superposition of states --- starting from the 
set of separable states --- or by taking derivatives of curves lying on a variety
defining a class of entanglement --- starting again from the set of separable states. This second construction
amounts to building new states as limits of
 superpositions (Section \ref{tools}) and corresponds to exceptional states\cite{ST}. 
In our atlas, the subvarieties of the nullcone are all obtained by this second construction.

For higher dimensional varieties, our method fails to be exhaustive. 
The representation $\mathcal{H}=\CC^2\otimes\CC^2\otimes\CC^2\otimes \CC^2$ of $G$ is {\em tame},
which means  in particular that $G$ acts with finitely many orbits on the nullcone,
but the number of $G$-orbit in $\mathcal{H}$ is infinite.
This is already the case for the third secant variety. 
In this case our initial algorithm was not able see 
 the variety $\text{Osc}(X)$ (again this variety generalize the W-states).
We had to modify our algorithm by geometric arguments to detect this variety.
 
This suggests how to further investigate the geometry of the four qubit systems. 
If we relax the condition on
 the vanishing of the invariants --- the conditions which defined here the nullcone and the third secant ---
 our algorithm will detect many more orbits, but  may also miss some important geometric objects. For example the study the 
 entanglement of the
 states satisfying  $L=0$ will require a better geometric understanding of the $G$-varieties of $\PP(\mathcal{H})$ to select
 the invariants for our algorithm. The choice of the conditions need to be discussed. 
In particular, we point out the role of the invariants of the quartic form
$$x_0^4-2B_{0000}x_0^3x_1+(B_{0000}^2+2L+4M)x_0^2x_1^2-4(B_{0000}(M+\frac12L)-D_{xy})x_0x_1^3+L^2x_1^4.$$
 This polynomial has two interesting properties: first, when one evaluates it on the $G_ {abcd}$ state \cite{VDMV}, its roots are $a^2,\,b^2,\,c^2$ and $d^2$; furthermore  its discriminant is $\Delta$. This work will be presented in an forthcoming paper\cite{HLT2}.\\ 
 
\noindent{\bf Acknowledgments}: This paper is partially supported by the PEPS-ICQ project 
COGIT (COmbinatoire et G\'eom\'etrie pour l'InTrication)
of the CNRS.

\appendix
\section{A complete system of polynomial covariants\label{AppCov}}

 The only covariant of degree $1$ is the ground form $A$.
\begin{equation}\begin{array}{cc}\mbox{Degree 2}&\mbox{Degree 3}\\
\begin{array}{|c|c|}
\hline \mbox{Symbol}&\mbox{Transvectant}\\
\hline
B_{0000}&\frac12(A,A)^{1111}\\
\hline B_{2200}&\frac12(A,A)^{0011}\\
B_{2020}&\frac12(A,A)^{0101}\\
B_{2002}&\frac12(A,A)^{0110}\\
B_{0220}&\frac12(A,A)^{1001}\\
B_{0202}&\frac12(A,A)^{1010}\\
B_{0022}&\frac12(A,A)^{1100}\\\hline
\end{array}\nonumber&\begin{array}{|c|c|}
\hline \mbox{Symbol}&\mbox{Transvectant}\\
\hline C^1_{1111}&(A,B_{2200})^{1100}+(A,B_{0022})^{0011}\\
 C^2_{1111}&(A,B_{0220})^{0110}+(A,B_{2002})^{1001}
\\\hline
 C_{3111}&\frac13\left((A,B_{2200})^{0100}+(A,B_{2020})^{0010}+(A,B_{2002})^{0001}\right)\\
 C_{1311}&\frac13\left((A,B_{2200})^{1000}+(A,B_{0220})^{0010}+(A,B_{0202})^{0001}\right)\\
 C_{1131}&\frac13\left((A,B_{2020})^{1000}+(A,B_{0220})^{0100}+(A,B_{0022})^{0001}\right)\\
 C_{1113}&\frac13\left((A,B_{2002})^{1000}+(A,B_{0202})^{0100}+(A,B_{0022})^{0010}\right)\\
 \hline
\end{array}\nonumber \end{array}
\end{equation}
\begin{center}Degree 4
\begin{equation}\begin{array}{cc}\begin{array}{|c|c|}\hline
\mbox{Symbol}&\mbox{Transvectant}\\\hline
D^1_{0000}&(A,C^1_{1111})^{1111}\\
D^2_{0000}&(A,C^2_{1,1,1,1})^{1111}\\\hline
D_{2200}&(A,C_{1111}^1)^{0011}\\
D_{2020}&(A,C^1_{1111})^{0101}\\
D_{2002}&(A,C^1_{1111})^{0110}\\
D_{0220}&(A,C^1_{1111})^{1001}\\
D_{0202}&(A,C_{1111}^1)^{1010}\\
D_{0022}&(A,C_{1111})^{1100}\\\hline
\end{array}&\begin{array}{|c|c|}\hline
\mbox{Symbol}&\mbox{Transvectant}\\\hline
D_{4000}&(A,C_{3111})^{0111}\\
D_{0400}&(A,C_{1311})^{1011}\\
D_{0040}&(A,C_{1131})^{1101}\\
D_{0004}&(A,C_{1113})^{1110}\\\hline
D^1_{2220}&(A,C_{1111}^1)^{0001}\\
D^2_{2220}&(A,C^2_{1111})^{0001}\\
D^1_{2202}&(A,C_{1111}^1)^{0010}\\
D^2_{2202}&(A,C_{1111}^2)^{0010}\\
D^1_{2022}&(A,C_{1111}^1)^{0100}\\
D^2_{2022}&(A,C^2_{1111})^{0100}\\
D^1_{0222}&(A,C_{1111}^1)^{1000}\\
D^2_{0222}&(A,C_{1111}^2)^{1000}\\\hline
\end{array}\end{array}\nonumber\end{equation}\end{center}
\begin{center} Degree 5
\begin{equation}\begin{array}{|c|c|}\hline
\mbox{Symbol}&\mbox{Transvectant}\\\hline
E_{1111}&(A,D_{2200})^{1100}\\\hline
E^1_{3111}&(A,D_{2200})^{0100}+(A,D_{2020})^{0010}+(A,D_{2002})^{0001}\\
E^2_{3111}&(A,D_{2220}^1)^{0110}-(A,D_{2022}^1)^{0011}+(A,D_{2202}^1)^{0101}\\
E^3_{3111}&(A,D_{2220}^2)^{0110}-(A,D_{2022}^2)^{0011}+(A,D_{2202}^2)^{0101}\\
E^1_{1311}&(A,D_{2200})^{1000}+(A,D_{0220})^{0010}+(A,D_{0202})^{0001}\\
E^2_{1311}&(A,D_{2220}^1)^{1010}-(A,D_{2202}^1)^{1001}+(A,D_{0222}^1)^{0011}\\
E^3_{1311}&f,D_{2220}^2)^{1010}-(A,D_{2202}^2)^{1001}+(A,D_{0222}^2)^{0011}\\
E^1_{1131}&(A,D_{2020})^{1000}+(A,D_{0220})^{0100}+(A,D_{0022})^{0001}\\
E^2_{1131}&(A,D_{2220}^1)^{1100}-(A,D_{2022}^1)^{1001}+(A,D_{0222}^1)^{0101}\\
E^3_{1131}&(A,D_{2220}^2)^{1100}-(A,D_{2022}^2)^{1001}+(A,D_{0222}^2)^{0101}\\
E^1_{1113}&(A,D_{2002})^{1000}+(A,D_{0202})^{0100}+(A,D_{0022})^{0010}\\
E^2_{1113}&(A,D_{2202}^1)^{1100}-(A,D_{2022}^1)^{1010}+(A,D_{0222}^1)^{0110}\\
E^3_{1113}&(A,D_{2202}^2)^{1100}-(A,D_{2022}^2)^{1010}+(A,D_{0222}^2)^{0110}\\\hline
\end{array}
\nonumber\end{equation}\end{center}
\begin{center}Degree 6
\begin{equation}\begin{array}{cc}\begin{array}{|c|c|}\hline
\mbox{Symbol}&\mbox{Transvectant}\\\hline
F_{0000}&(A,E_{1111})^{1111}\\\hline
F_{2200}&(A,E^1_{1111})^{0011}\\
F_{2020}&(A,E_{1111})^{0101}\\
F_{2002}&(A,E^1_{1111})^{0110}\\
F_{0220}&(A,E^1_{1111})^{1001}\\
F_{0202}&(A,E^3_{1111})^{1010}\\
F_{0022}&(A,E^1_{1111})^{1100}\\\hline
F_{2220}^1&(A,E^1_{1311})^{0101}-(A,E^1_{3111})^{1001})+(A,E^1_{1131})^{0011}\\
F_{2220}^2&(A,E^1_{1311})^{0101}+(A,E^1_{3111})^{1001})-(A,E^1_{1131})^{0011}\\
F_{2202}^1&(A,E^1_{1311})^{0110}-(A,E^1_{3111})^{1010})+(A,E^1_{1113})^{0011}\\
F_{2202}^2&(A,E^1_{1311})^{0110}+(A,E^1_{3111})^{1010})-(A,E^1_{1113})^{0011}\\
F_{2022}^1&(A,E^1_{3111})^{1100}-(A,E^1_{1131})^{0110})+(A,E^1_{1113})^{0101}\\
F_{2022}^2&(A,E^1_{3111})^{1100}+(A,E^1_{1131})^{0110})-(A,E^1_{1113})^{0101}\\
F_{0222}^1&(A,E^1_{1311})^{1100}-(A,E^1_{1131})^{1010})+(A,E^1_{1113})^{1001}\\
F_{0222}^2&(A,E^1_{1311})^{1100}+(A,E^1_{1131})^{1010})-(A,E^1_{1113})^{1001}\\
\hline\end{array}&
\begin{array}{|c|c|}\hline
\mbox{Symbol}&\mbox{Transvectant}\\\hline
F_{4200}&(A,E^1_{3111})^{0011}\\
F_{4020}&(A,E^1_{3111})^{0101}\\
F_{4002}&(A,E^1_{3111})^{0110}\\
F_{0420}&(A,E^1_{1311})^{1001}\\
F_{0402}&(A,E^1_{1311})^{1010}\\
F_{0042}&(A,E^1_{1131})^{1100}\\
F_{2400}&(A,E^1_{1311})^{0011}\\
F_{2040}&(A,E^1_{1131})^{0101}\\
F_{2004}&(A,E^1_{1113})^{0110}\\
F_{0240}&(A,E^1_{1131})^{1001}\\
F_{0204}&(A,E^1_{1113})^{1010}\\
F_{0024}&(A,E^1_{1113})^{1100}\\\hline
\end{array}\end{array}\nonumber\end{equation}
\end{center}
\begin{center}Degree 7
\begin{equation}\begin{array}{cc}\begin{array}{|c|c|}\hline
\mbox{Symbol}&\mbox{Transvectant}\\\hline
G^1_{3111}&(A,F_{4200})^{1100}\\
G^2_{3111}&(A,F_{4020})^{1010}\\
G^3_{3111}&(A,F_{4002})^{1001}\\
G^1_{1311}&(A,F_{2400})^{110}\\
G^2_{1311}&(A,F_{0420})^{0110}\\
G^3_{1311}&(A,F_{0402})^{0101}\\
G^1_{1131}&(A,F_{2040})^{1010}\\
G^2_{1131}&(A,F_{0240})^{0110}\\
G^3_{1131}&(A,F_{0024})^{0011}\\
G^1_{1113}&(A,F_{2004})^{1001}\\
G^2_{1113}&(A,F_{0204})^{0101}\\
G^3_{1113}&(A,F_{0024})^{0011}
\\\hline\end{array}&
\begin{array}{|c|c|}\hline
\mbox{Symbol}&\mbox{Transvectant}\\\hline
G_{5111}&(A,F_{4002})^{0001}+(A,F_{4020})^{0010}+(A,F_{4200})^{0100}\\
G_{1511}&(A,F_{0402})^{0001}+(A,F_{0420})^{0010}+(A,F_{2400})^{1000}\\
G_{1151}&(A,F_{0042})^{0001}+(A,F_{0240})^{0100}+(A,F_{2040})^{1000}\\
G_{1115}&(A,F_{0204})^{0100}+(A,F_{0024})^{0010}+(A,F_{2004})^{1000}\\\hline
G_{3311}&(A,F_{2400})^{0100}\\
G_{3131}&(A,F_{2040})^{0010}\\
G_{3113}&(A,F_{2004})^{0001}\\
G_{1331}&(A,F_{0240})^{0010}\\
G_{1313}&(A,F_{0204})^{0001}\\
G_{1133}&(A,F_{0024})^{0001}\\\hline
\end{array}\end{array}\nonumber\nonumber
\end{equation}
\end{center}
\begin{center} Degree 8
\begin{equation}\begin{array}{cc}\begin{array}{|c|c|}\hline
\mbox{Symbol}&\mbox{Transvectant}\\\hline
H_{4000}&(A,G_{3111})^{0111}\\
H_{0400}&(A,G_{1311})^{1011}\\
H_{0040}&(A,G_{1131})^{1101}\\
H_{0004}&(A,G_{1113})^{1110}
\\\hline
H_{2220}^1&(A,G_{1311}^1)^{0101}+(A,G_{3111}^1)^{1001}+(A,G_{1131}^1)^{0011}\\
H_{2220}^2&(A,G_{1311}^2)^{0101}+(A,G_{3111}^2)^{1001}+(A,G_{1131}^2)^{0011}\\
H_{2202}^1&(A,G_{1311}^1)^{0110}+(A,G_{3111}^1)^{1010}+(A,G_{1113}^1)^{0011}\\
H_{2202}^2&(A,G_{1311}^2)^{0110}+(A,G_{3111}^2)^{1010}+(A,G_{1113}^2)^{0011}\\
H_{2022}^1&(A,G_{3111}^1)^{1100}+(A,G_{1131}^1)^{0110}+(A,G_{1113}^1)^{0101}\\
H_{2022}^2&(A,G_{3111}^2)^{1100}+(A,G_{1131}^2)^{0110}+(A,G_{1113}^2)^{0101}\\
H_{0222}^1&(A,G_{1311}^1)^{1100}+(A,G_{1131}^1)^{1010}+(A,G_{1113}^1)^{1001}\\
H^2_{0222}&(A,G_{1311}^2)^{1100}+(A,G_{1131}^2)^{1010}+(A,G_{1113}^2)^{1001}\\\hline\end{array}&
\begin{array}{|c|c|}\hline
\mbox{Symbol}&\mbox{Transvectant}\\\hline
H_{4200}&(A,G_{5111})^{1011}\\
H_{4020}&(A,G_{5111})^{1101}\\
H_{4002}&(A,G_{5111})^{1110}\\
H_{0420}&(A,G_{1511})^{1101}\\
H_{0402}&(A,G_{1511})^{1110}\\
H_{0042}&(A,G_{1151})^{1110}\\
H_{2400}&(A,G_{1511}^1)^{0111}\\
H_{2040}&(A,G_{1151})^{0111}\\
H_{2004}&(A,G_{1115}^1)^{0111}\\
H_{0240}&(A,G_{1151})^{1011}\\
H_{0204}&(A,G_{1115})^{1011}\\
H_{0024}&(A,G_{1115}^1)^{1101}\\\hline
\end{array}\end{array}\nonumber
\end{equation}\end{center}
\begin{center} Degree 9
\begin{equation}\begin{array}{cc}\begin{array}{|c|c|}\hline
\mbox{Symbol}&\mbox{Transvectant}\\\hline
I_{3111}&(A,H_{4020})^{1010}+(A,H_{4200})^{1100}+(A,H_{4002})^{1001}\\
I_{1311}&(A,H_{0420})^{0110}+(A,H_{2400})^{1100}+(A,H_{0402})^{0101}\\
I_{1131}&(A,H_{0240})^{0110}+(A,H_{2040})^{1010}+(A,H_{0042})^{0011}\\
I_{1113}&(A,H_{0204})^{0101}+(A,H_{2004})^{1001}+(A,H_{0024})^{0011}\\\hline
I_{5111}^1&(A,H_{4020})^{0010}+(A,H_{4200})^{0100}+(A,H_{4002})^{0001}\\
I_{5111}^2&(A,H_{4020})^{0010}-(A,H_{4200})^{0100}+(A,H_{4002})^{0001}\\
I_{1511}^1&(A,H_{0420})^{0010}+(A,H_{2400})^{1000}+(A,H_{4002})^{0001}\\
I_{1511}^2&(A,H_{0420})^{0010}-(A,H_{2400})^{1000}+(A,H_{4002})^{0001}\\
I_{1151}^1&(A,H_{0240})^{0100}+(A,H_{2040})^{1000}+(A,H_{0042})^{0001}\\
I_{1151}^2&(A,H_{0240})^{0100}-(A,H_{2040})^{1000}+(A,H_{0042})^{0001}\\
I_{1115}^1&(A,H_{0204})^{0100}+(A,H_{2004})^{1000}+(A,H_{0024})^{0010}\\
I_{1115}^2&(A,H_{0204})^{0100}-(A,H_{2004})^{1000}+(A,H_{0024})^{0010}\\\hline
\end{array}&
\begin{array}{|c|c|}\hline
\mbox{Symbol}&\mbox{Transvectant}\\\hline
I_{3311}^1&(A,H_{2220}^1)^{0010}+(A,H_{2202}^1)^{0001}\\
I_{3311}^2&(A,H_{2220}^2)^{0010}+(A,H_{2202}^2)^{0001}\\
I_{3131}^1&(A,H_{2220}^1)^{0100}+(A,H_{2022}^1)^{0001}\\
I_{3131}^2&(A,H_{2220}^2)^{0100}+(A,H_{2022}^2)^{0001}\\
I_{3113}^1&(A,H_{2202}^1)^{0100}+(A,H_{2022}^1)^{0010}\\
I_{3113}^2&(A,H_{2202}^2)^{0100}+(A,H_{2022}^2)^{0010}\\
I_{1331}^1&(A,H_{2220}^1)^{1000}+(A,H_{0222}^1)^{0001}\\
I_{1331}^2&((A,H_{2202}^2)^{0100}+(A,H_{2022}^2)^{0010}\\
I_{1313}^1&(A,H_{0222}^1)^{0010}+(A,H_{2202}^1)^{1000}\\
I_{1313}^2&(A,H_{0222}^2)^{0010}+(A,H_{2202}^2)^{1000}\\
I_{1133}^1&(A,H_{0222}^1)^{010}+(A,H_{2022}^1)^{1000}\\
I_{1133}^2&(A,H_{0222}^2)^{010}+(A,H_{2022}^2)^{1000}\\\hline
\end{array}\end{array}\nonumber
\end{equation}\end{center}

\begin{equation}\begin{array}{cc}\mbox{Degree 10}&\\\begin{array}{|c|c|}\hline
\mbox{Symbol}&\mbox{Transvectant}\\\hline
J_{4200}&(A,I_{5111})^{1011}\\
J_{4020}&(A,I_{5111})^{1101}\\
J_{4002}&(A,I_{5111})^{1110}\\
J_{0420}&(A,I_{1511})^{1101}\\
J_{0402}&(A,I_{1511})^{1110}\\
J_{0042}&(A,I_{1151})^{1110}\\
J_{2400}&(A,I_{1511})^{0111}\\
J_{2040}&(A,I_{1151})^{0111}\\
J_{2004}&(A,I_{1115})^{0111}\\
J_{0240}&(A,I_{1151})^{1011}\\
J_{0204}&(A,I_{1115})^{1011}\\
J_{0024}&(A,I_{1115})^{1101}\\\hline
\end{array}&\begin{array}{c}\mbox{Degree 11}\\
\begin{array}{|c|c|}\hline
\mbox{Symbol}&\mbox{Transvectant}\\\hline
K_{3311}&=(A,J_{4200})^{1000}-(A,J_{2400})^{0100}\\
K_{3131}&=(A,J_{4020})^{1000}-(A,J_{2040})^{0010}\\
K_{3113}&=(A,J_{4002})^{1000}-(A,J_{2004})^{0001}\\
K_{1331}&=(A,J_{0420})^{0100}-(A,J_{0240})^{0010}\\
K_{1313}&=(A,J_{0402})^{0100}-(A,J_{0204})^{0001}\\
K_{1133}&=(A,J_{0042})^{0010}-(A,J_{0024})^{0001}\\\hline\hline
K_{5111}&=(A,J_{4200})^{0100}-(A,J_{4020})^{0010}+(A,J_{4002})^{0001}\\
K_{1511}&=(A,J_{2400})^{1000}-(A,J_{0420})^{0010}+(A,J_{0402})^{0001}\\
K_{1151}&=(A,J_{2040})^{1000}-(A,J_{0240})^{0100}+(A,J_{0042})^{0001}\\
K_{1115}&=(A,J_{2004})^{1000}-(A,J_{0204})^{0110}+(A,J_{0024})^{0010}\\\hline
\end{array}\\
\mbox{Degree 12} \\
\begin{array}{|c|c|}\hline
\mbox{Symbol}&\mbox{Transvectant}\\\hline
L_{6000}&=(A,K_{5111})^{0111}\\
L_{0600}&=(A,K_{1511})^{1011}\\
L_{0060}&=(A,K_{1151})^{1101}\\
L_{0006}&=(A,K_{1115})^{1110}\\\hline\end{array}\end{array}\end{array}\nonumber\end{equation}

\section{Nullcone: evaluation of $T$\label{NulCT}}
Evaluating $T$ on the different forms we find:
{\footnotesize\[
\begin{array}{|c|c|c|c|}
\hline
65511&65218&65271&65247\\\hline
\begin{array}{c}
1\\ 
1, 1, 1, 1, 1, 1\\ 
1, 1, 1, 1\\ 
1, 1, 1, 1\\ 
1, 1, 1, 1, 1, 1\\
1, 1, 1, 1\\ 
0, 0, 0, 1
\end{array}&
\begin{array}{c}1\\ 
	1, 1, 1, 1, 1, 1\\
	 1, 1, 1, 1\\
	 1, 1, 1, 1\\
	 1, 1, 1, 1, 1, 1\\
1, 1, 1, 1\\ 
1, 0, 0, 0\end{array}
&
\begin{array}{c}
1\\ 
	1, 1, 1, 1, 1, 1\\
	 1, 1, 1, 1\\
	  1, 1, 1, 1\\
	  1, 1, 1, 1, 1, 1\\
1, 1, 1, 1\\ 
0, 0, 1, 0\end{array}
&
\begin{array}{c}
1\\ 
	1, 1, 1, 1, 1, 1\\
	 1, 1, 1, 1\\
	  1, 1, 1, 1\\
	  1, 1, 1, 1, 1, 1\\
1, 1, 1, 1\\ 
0, 1, 0, 0\end{array}
\\\hline
\end{array}
\]
}
{\footnotesize\[
\begin{array}{|c|c|c|c|}
\hline
65508&64762&65506&65482\\\hline
\begin{array}{c}
1\\ 
1, 1, 1, 1, 1, 1\\ 
1, 1, 1, 1\\ 
0, 1, 1, 1\\ 
0, 0, 0, 1, 1, 1\\
0, 0, 0, 1\\ 
0, 0, 0, 0
\end{array}&
\begin{array}{c}1\\ 
	1, 1, 1, 1, 1, 1\\
	 1, 1, 1, 1\\
	 1, 1, 1, 0\\
	 1, 1, 0, 1, 0, 0\\
1, 0, 0, 0\\ 
0, 0, 0, 0\end{array}
&
\begin{array}{c}
1\\ 
	1, 1, 1, 1, 1, 1\\
	 1, 1, 1, 1\\
	  1, 0, 1, 1\\
	  0, 1, 1, 0, 0, 1\\
0, 0, 1, 0\\ 
0, 0, 0, 0\end{array}
&
\begin{array}{c}
1\\ 
	1, 1, 1, 1, 1, 1\\
	 1, 1, 1, 1\\
	  1, 1, 0, 1\\
	  1, 0, 1, 0, 1, 0\\
0, 1, 0, 0\\ 
0, 0, 0, 0\end{array}
\\\hline
\end{array}
\]
}

{\footnotesize\[
\begin{array}{|c|c|c|c|c|c|}
\hline
64700&65041&65075&61158&65109&64218\\\hline
\begin{array}{c}
1\\ 
1, 1, 1, 1, 1, 1\\ 
1, 1, 1, 1\\ 
0, 1, 1, 0\\ 
0, 0, 0, 1, 0, 0\\
0, 0, 0, 0\\ 
0, 0, 0, 0
\end{array}
&
\begin{array}{c}
1\\ 
1, 1, 1, 1, 1, 1\\ 
1, 1, 1, 1\\ 
0, 0, 1, 1\\ 
0, 0, 0, 0, 0, 1\\
0, 0, 0, 0\\ 
0, 0, 0, 0
\end{array}&
\begin{array}{c}1\\ 
	1, 1, 1, 1, 1, 1\\
	 1, 1, 1, 1\\
	 0, 1, 0, 1\\
	 0, 0, 0, 0, 1, 0\\
0, 0, 0, 0\\ 
0, 0, 0, 0\end{array}
&
\begin{array}{c}
1\\ 
	1, 1, 1, 1, 1, 1\\
	 1, 1, 1, 1\\
	  1, 1, 0, 0\\
	  1, 0, 0, 0, 0, 0\\
0, 0, 0, 0\\ 
0, 0, 0, 0\end{array}
&
\begin{array}{c}
1\\ 
	1, 1, 1, 1, 1, 1\\
	 1, 1, 1, 1\\
	  1, 0, 0, 1\\
	  0, 0, 1, 0, 0, 0\\
0, 0, 0, 0\\ 
0, 0, 0, 0\end{array}
&
\begin{array}{c}
1\\ 
	1, 1, 1, 1, 1, 1\\
	 1, 1, 1, 1\\
	  1, 0, 1, 0\\
	  0, 1, 0, 0, 0, 0\\
0, 0, 0, 0\\ 
0, 0, 0, 0\end{array}
\\\hline
\end{array}
\]
}

{\footnotesize\[
\begin{array}{|c|c|c|c|}
\hline
65530&65518&65532&65278\\\hline
\begin{array}{c}
1\\ 
1, 0, 0, 1, 1, 0\\ 
0, 1, 0, 0\\ 
0, 1, 0, 0\\ 
0, 0, 0, 0, 0, 0\\
0, 0, 0, 0\\ 
0, 0, 0, 0
\end{array}&
\begin{array}{c}1\\ 
	0, 1, 0, 1, 0, 1\\
	 0, 0, 1, 0\\
	 0, 0, 1, 0\\
	 0, 0, 0, 0, 0, 0\\
0, 0, 0, 0\\ 
0, 0, 0, 0\end{array}
&
\begin{array}{c}
1\\ 
	1, 1, 1, 0, 0, 0\\
	 1, 0, 0, 0\\
	  1, 0, 0, 0\\
	  0, 0, 0, 0, 0, 0\\
0, 0, 0, 0\\ 
0, 0, 0, 0\end{array}
&
\begin{array}{c}
1\\ 
	0, 0, 1, 0, 1, 1\\
	 0, 0, 0, 1\\
	  0, 0, 0, 1\\
	  0, 0, 0, 0, 0, 0\\
0, 0, 0, 0\\ 
0, 0, 0, 0\end{array}
\\\hline
\end{array}
\]
}

{\footnotesize\[
\begin{array}{|c|}
\hline
59520\\\hline
\begin{array}{c}
1\\ 
1, 1, 1, 1, 1, 1\\ 
1, 1, 1, 1\\ 
0, 0, 0, 0\\ 
0, 0, 0, 0, 0, 0\\
0, 0, 0, 0\\ 
0, 0, 0, 0
\end{array}\\\hline\end{array}\]
}

{\footnotesize\[
\begin{array}{|c|c|c|c|}
\hline
64160&61064&64704&59624\\\hline
\begin{array}{c}
1\\ 
1, 0, 0, 1, 1, 0\\ 
0, 1, 0, 0\\ 
0, 0, 0, 0\\ 
0, 0, 0, 0, 0, 0\\
0, 0, 0, 0\\ 
0, 0, 0, 0
\end{array}&
\begin{array}{c}1\\ 
	0, 1, 0, 1, 0, 1\\
	 0, 0, 1, 0\\
	 0, 0, 0, 0\\
	 0, 0, 0, 0, 0, 0\\
0, 0, 0, 0\\ 
0, 0, 0, 0\end{array}
&
\begin{array}{c}
1\\ 
	1, 1, 1, 0, 0, 0\\
	 1, 0, 0, 0\\
	  0, 0, 0, 0\\
	  0, 0, 0, 0, 0, 0\\
0, 0, 0, 0\\ 
0, 0, 0, 0\end{array}
&
\begin{array}{c}
1\\ 
	0, 0, 1, 0, 1, 1\\
	 0, 0, 0, 1\\
	  0, 0, 0, 0\\
	  0, 0, 0, 0, 0, 0\\
0, 0, 0, 0\\ 
0, 0, 0, 0\end{array}
\\\hline
\end{array}
\]
}

{\footnotesize\[
\begin{array}{|c|c|c|c|c|c|}
\hline
65520&65484&65450&64764&64250&61166\\\hline
\begin{array}{c}
1\\ 
1, 0, 0, 0, 0, 0\\ 
0, 0, 0, 0\\ 
0, 0, 0, 0\\ 
0, 0, 0, 0, 0, 0\\
0, 0, 0, 0\\ 
0, 0, 0, 0
\end{array}
&
\begin{array}{c}
1\\ 
0, 1, 0, 0, 0, 0\\ 
0, 0, 0, 0\\ 
0, 0, 0, 0\\ 
0, 0, 0, 0, 0, 0\\
0, 0, 0, 0\\ 
0, 0, 0, 0
\end{array}&
\begin{array}{c}1\\ 
	0, 0, 0, 1, 0, 0\\
	 0, 0, 0, 0\\
	 0, 0, 0, 0\\
	 0, 0, 0, 0, 0, 0\\
0, 0, 0, 0\\ 
0, 0, 0, 0\end{array}
&
\begin{array}{c}
1\\ 
	0, 0, 1, 0, 0, 0\\
	 0, 0, 0, 0\\
	  0, 0, 0, 0\\
	  0, 0, 0, 0, 0, 0\\
0, 0, 0, 0\\ 
0, 0, 0, 0\end{array}
&
\begin{array}{c}
1\\ 
	0, 0, 0, 0, 1, 0\\
	 0, 0, 0, 0\\
	  0, 0, 0, 0\\
	  0, 0, 0, 0, 0, 0\\
0, 0, 0, 0\\ 
0, 0, 0, 0\end{array}
&
\begin{array}{c}
1\\ 
	0, 0, 0, 0, 0, 1\\
	 0, 0, 0, 0\\
	  0, 0, 0, 0\\
	  0, 0, 0, 0, 0, 0\\
0, 0, 0, 0\\ 
0, 0, 0, 0\end{array}
\\\hline
\end{array}
\]
}

{\footnotesize\[
\begin{array}{|c|}
\hline
65535\\\hline
\begin{array}{c}
1\\ 
0, 0, 0, 0, 0, 0\\ 
0, 0, 0, 0\\ 
0, 0, 0, 0\\ 
0, 0, 0, 0, 0, 0\\
0, 0, 0, 0\\ 
0, 0, 0, 0
\end{array}\\\hline\end{array}\]
}

{\footnotesize\[
\begin{array}{|c|}
\hline
0\\\hline
\begin{array}{c}
0\\ 
0, 0, 0, 0, 0, 0\\ 
0, 0, 0, 0\\ 
0, 0, 0, 0\\ 
0, 0, 0, 0, 0, 0\\
0, 0, 0, 0\\ 
0, 0, 0, 0
\end{array}\\\hline\end{array}\]
}

\section{Geometric inclusions}\label{inclusion}
We restate the inclusion diagrams of Figures \ref{GR''4}, \ref{GR''3}, \ref{GR''2} and \ref{GR''1} with our geometric interpretations.

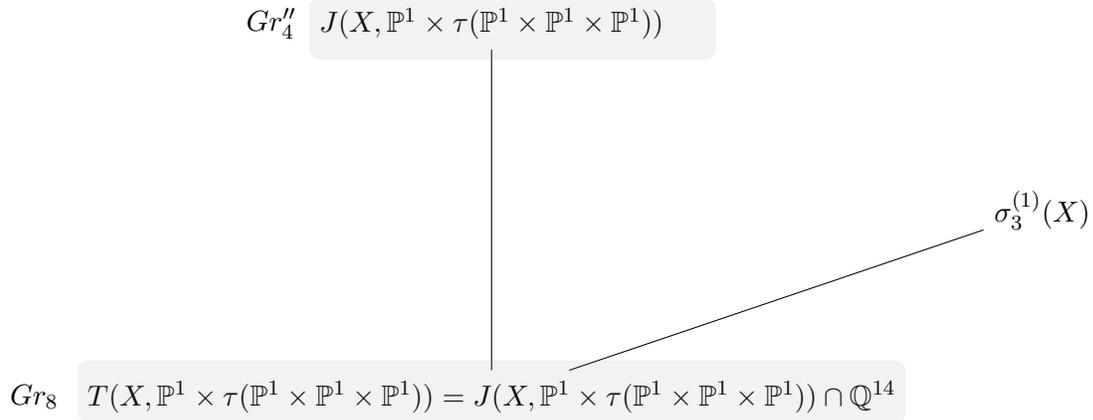
\begin{figure}[t]
\begin{center}
\begin{tikzpicture}
 \matrix (mat) [matrix of nodes,ampersand replacement=\&,
     row sep=50pt,column sep=10pt, left delimiter={.}, right delimiter={.},
      nodes={minimum height=0.5cm,minimum width=1.5cm }]{
$J(X,\PP^1\times\tau(\PP^1\times\PP^1\times\PP^1))$\&  \\
           \&   \&   \&  $\sigma_3^{(1)}(X)$\\ 
 $T(X,\PP^1\times\tau(\PP^1\times\PP^1\times\PP^1))=J(X,\PP^1\times\tau(\PP^1\times\PP^1\times\PP^1))\cap \QQ^{14}$\& \\};
\node (p65513) [rectangle,rounded corners, inner sep=0pt, fit= (mat-1-1) (mat-1-1)] {};
\node (p59777) [rectangle,rounded corners, inner sep=0pt, fit= (mat-2-4) (mat-2-4)] {};
\node (p65511) [rectangle,rounded corners, inner sep=0pt, fit= (mat-3-1) (mat-3-1)] {};
\draw (p65511)--(p65513);\draw (p65511)--(p59777);
 \node (Gr''4) [rectangle,rounded corners,fill=gray,opacity=0.1, inner sep=0pt, fit= (mat-1-1) (mat-1-4)] {};
 \node (Gr8) [rectangle,rounded corners,fill=gray,opacity=0.1, inner sep=0pt, fit= (mat-3-1) (mat-3-4)] {};
\node [left=70pt] at (p65513)  {$Gr''_4$}; 
\node [left=160pt] at (p65511)  {$Gr_8$}; 
\end{tikzpicture}
\end{center}
\caption{Geometric interpretation of the inclusions of the stratas $Gr''_4$, $59777$ and $Gr_8$\label{GR''4bis}}
\end{figure}

Figure \ref{GR''4bis} is the geometric analogue of Figure \ref{GR''4}. The inclusions between varieties 
of groups $Gr''_4$ and $Gr_8$ are 
just the natural inclusions of the tangential and join varieties. The variety $\sigma^{(1)}_3(X)=\sigma_3(X)\cap \QQ^{14}$ 
contains the varieties of type 
$Gr_8$ which are also given by the intersection of varieties of $Gr_4''$ with the hypersurface $\QQ^{14}$.

\begin{figure}[h]
\begin{center}
\begin{tikzpicture}
 \matrix (mat) [matrix of nodes,ampersand replacement=\&,
     row sep=50pt,column sep=10pt, left delimiter={.}, right delimiter={.},
      nodes={minimum height=0.5cm,minimum width=1.5cm }]{
$J(\PP^3\times\PP^1\times\PP^1,\PP^1\times\PP^3\times\PP^1)$\&  \\
           \&   \&   \&  $\text{Osc}'(X)$\\ 
 $Z_3(X)$\& $Z_4(X)$ \\};
\node (p65513) [rectangle,rounded corners, inner sep=0pt, fit= (mat-1-1) (mat-1-1)] {};
\node (p59777) [rectangle,rounded corners, inner sep=0pt, fit= (mat-2-4) (mat-2-4)] {};
\node (p65511) [rectangle,rounded corners, inner sep=0pt, fit= (mat-3-1) (mat-3-1)] {};
\node (p64762) [rectangle,rounded corners, inner sep=0pt, fit= (mat-3-2) (mat-3-2)] {};
\draw (p65511)--(p65513);\draw (p65511)--(p59777);
\draw (p64762)--(p65513);\draw (p64762)--(p59777);
 \node (Gr''4) [rectangle,rounded corners,fill=gray,opacity=0.1, inner sep=0pt, fit= (mat-1-1) (mat-1-4)] {};
 \node (Gr8) [rectangle,rounded corners,fill=gray,opacity=0.1, inner sep=0pt, fit= (mat-3-1) (mat-3-4)] {};
\node [left=100pt] at (p65513)  {$Gr''_4$}; 
\node [left=100pt] at (p65511)  {$Gr_8$}; 
\end{tikzpicture}
\end{center}
\caption{Geometric interpretation of the inclusions of the stratas $Gr''_3$, $59510$ and $Gr_7$\label{GR3''bis}}
\end{figure}
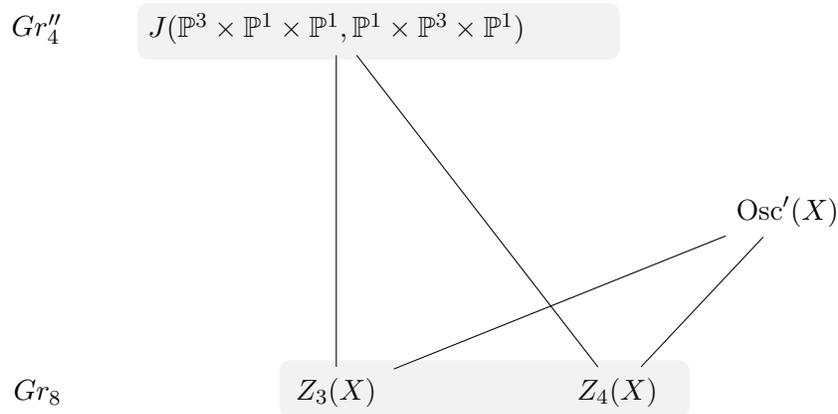

Let us make some comments on Figure \ref{GR''3} which we partially reproduced in Figure \ref{GR3''bis} with the corresponding varieties.

The normal form for a general element of $J(\PP^3\times\PP^1\times\PP^1,\PP^1\times\PP^3\times\PP^1)$ is 
\[\alpha\underbrace{|1001\rangle+|0101\rangle}_{\in \PP^3\times\PP^1\times\PP^1}+\beta\underbrace{|0000\rangle+|0110\rangle}_{\in \PP^1\times\PP^3\times\PP^1}.\]
If we look at the normal form of $Z_4(X)$ it naturally belongs to $J(\PP^3\times\PP^1\times\PP^1,\PP^1\times\PP^3\times\PP^1)$
\[\underbrace{|1000\rangle+|0100\rangle}_{\in \PP^3\times\PP^1\times\PP^1}+\underbrace{|0101\rangle+|0011\rangle}_{\in \PP^1\times\PP^3\times\PP^1}\]
Similarly a normal form for $Z_3(X)$ is 
\[\underbrace{|1000\rangle+|0001\rangle}_{\in\PP^1\times\PP^3\times\PP^1}+\underbrace{|1010\rangle+|0110\rangle}_{\in \PP^3\times\PP^1\times\PP^1}\]
proving $Z_3(X)\subset J(\PP^3\times\PP^1\times\PP^1,\PP^1\times\PP^3\times\PP^1)$.

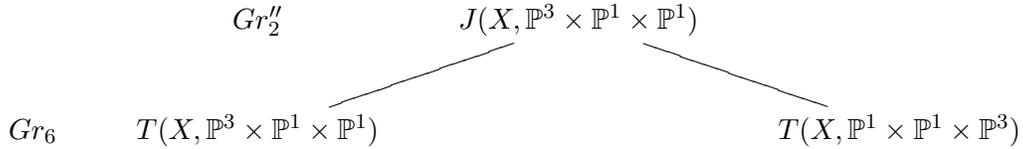
\begin{figure}[h]
 \[\xymatrix{&Gr_2'' & J(X,\PP^3\times\PP^1\times\PP^1)\incl[dr]\incl[dl] & \\
 Gr_6& T(X,\PP^3\times\PP^1\times\PP^1) & & T(X,\PP^1\times\PP^1\times\PP^3)\\
 }\]
 \caption{Geometric interpretation of the inclusions of stratas $Gr_2''$ and $Gr_6$}\label{GR''2bis}
\end{figure}

In Figure \ref{GR''2bis} we give the geometric version of Figure \ref{GR''2}. Again the inclusions $Gr_6\subset Gr_2''$ can be read
as the natural inclusion of tangential varieties and joins. Here we should notice that 
$J(X,\PP^3\times\PP^1\times\PP^1)=J(X,\PP^1\times\PP^1\times\PP^3)$.

\begin{figure}[t]
\begin{center}
\begin{tikzpicture}
 \matrix (mat) [matrix of nodes,ampersand replacement=\&,
     row sep=50pt,column sep=10pt, left delimiter={.}, right delimiter={.},
      nodes={minimum height=0.5cm,minimum width=1.5cm }]{
 $\sigma(X)$\&  \\
      \&      \&      \&       \& $\tau(X)$\\
 $\sigma(\PP^1\times\underline{\PP^1}\times\PP^1\times\PP^1)\times\PP^1$\& $\sigma(\PP^1\times\PP^1\times\underline{\PP^1}\times\PP^1)\times\PP^1$\& $\PP^1\times\PP^7$\& $\PP^7\times\PP^1$\&\\};
\node (65534) [rectangle,rounded corners, inner sep=0pt, fit= (mat-1-1) (mat-1-1)] {};
\node (59520) [rectangle,rounded corners, inner sep=0pt, fit= (mat-2-5) (mat-2-5)] {};
\node (65530) [rectangle,rounded corners, inner sep=0pt, fit= (mat-3-1) (mat-3-1)] {};
\node (65518) [rectangle,rounded corners, inner sep=0pt, fit= (mat-3-2) (mat-3-2)] {};
\node (65532) [rectangle,rounded corners, inner sep=0pt, fit= (mat-3-3) (mat-3-3)] {};
\node (65278) [rectangle,rounded corners, inner sep=0pt, fit= (mat-3-4) (mat-3-4)] {};

\draw (65534)--(59520);
\draw (65534)--(65278);
\draw (65534)--(65530);
\draw (65534)--(65518);
\draw (65534)--(65532);

 \node (Gr''1) [rectangle,rounded corners,fill=gray,opacity=0.1, inner sep=0pt, fit= (mat-1-1) (mat-1-1)] {};
 \node (Gr5) [rectangle,rounded corners,fill=gray,opacity=0.1, inner sep=0pt, fit= (mat-2-5) (mat-2-5)] {};
 \node (Gr4) [rectangle,rounded corners,fill=gray,opacity=0.1, inner sep=0pt, fit= (mat-3-1) (mat-3-4)] {};
\node [left=20pt] at (65534)  {$Gr''_1$}; 
\node [left=20pt] at (59520)  {$Gr_5$}; 
\node [left=70pt] at (65530)  {$Gr_4$}; 
\end{tikzpicture}
\end{center}
\caption{Geometric interpretations of the inclusions of stratas $Gr_1''$, $Gr_4$ and $Gr_5$}\label{GR''1bis}
\end{figure}
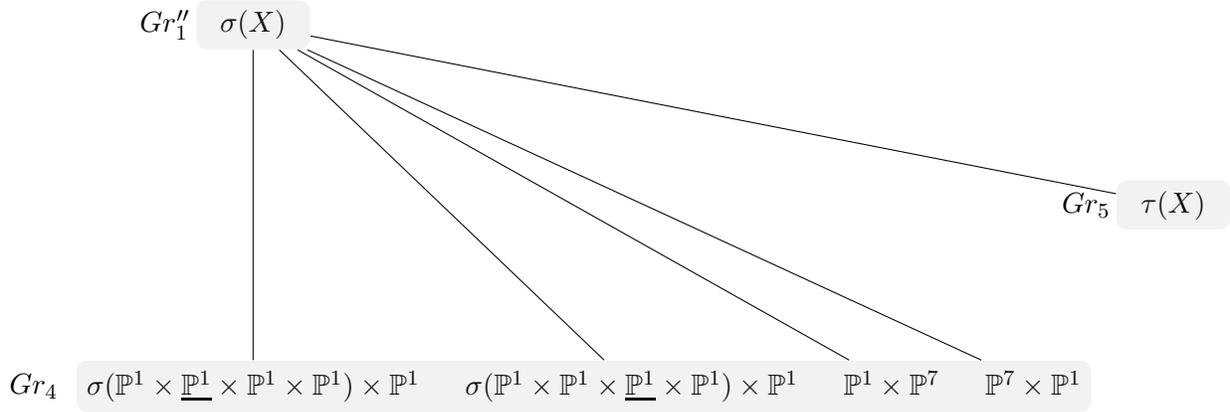
The inclusions $Gr_5\subset Gr_1''$ and $Gr_4\subset Gr_1''$ of Figure \ref{GR''1} are explained in Figure \ref{GR''1bis}. It corresponds 
to the natural inclusion $\tau(X)\subset\sigma(X)$ (the so-called ``onion'' structure between GHZ and W-states 
emphasized by Miyake for 
three qubits systems\cite{My,My2}) and the inclusions of partially entangled states ($Gr_4$) within the GHZ-states.

\end{document}